\documentclass[prd,aps,
nofootinbib,
floatfix,
superscriptaddress]{revtex4}
\usepackage{graphicx}
\usepackage{epsfig}
\usepackage{rotating}
\usepackage{amssymb}
\usepackage{subfigure}
\usepackage{dsfont}
\usepackage{psfrag}
\usepackage{amsmath,euscript,array,mathrsfs}

\newcommand{\SP}[1]{\begin{equation}\begin{split} #1
\end{split}\end{equation}}

\newcommand{\beq}{\begin{equation}}
\newcommand{\eeq}{\end{equation}}
\newcommand{\beqs}{\begin{eqnarray}}
\newcommand{\eeqs}{\end{eqnarray}}
\newcommand{\lsim}{\mathrel{\raisebox{-
.6ex}{$\stackrel{\textstyle<}{\sim}$}}}
\newcommand{\gsim}{\mathrel{\raisebox{-
.6ex}{$\stackrel{\textstyle>}{\sim}$}}}

\def\hbar{\hspace{0pt}\raisebox{1pt}{$-$} \hspace{-7pt} h}

\def\di{\mbox{d}}
\def\r{\rho}

\begin{document}
\title{Hyperscaling violation and Electroweak Symmetry Breaking}

\author{ Daniel Elander}
\affiliation{Department of Physics, Purdue University, 525 Northwestern Avenue, West Lafayette, IN 47907-2036, U.S.A.}
\author{Robert Lawrance}
\author{Maurizio Piai}
\affiliation{Department of Physics, College of Science, Swansea University,
Singleton Park, Swansea, Wales, UK}

\date{\today}


\begin{abstract}
We consider a class of simplified models of dynamical electroweak symmetry breaking 
built in terms of their five-dimensional weakly-coupled gravity duals, 
 in the spirit of bottom-up holography.
 The  sigma-model consists of two abelian gauge bosons
 and one real, non-charged scalar field coupled to gravity in five dimensions.
 The scalar potential is a simple exponential function of the scalar field.
The background metric resulting from solving the classical equations of motion
exhibits hyperscaling violation, at least  at  asymptotically large values of the radial direction.
We study  the spectrum of scalar composite states of the putative dual field theory 
by fluctuating the sigma-model scalars and gravity, and discuss in which cases we find a parametrically light scalar state in the spectrum.
We model the spontaneous breaking of the  (weakly coupled) 
gauge symmetry  to the diagonal subgroup by the choice of IR boundary conditions. We compute the mass spectrum  of spin-1 states,
and  the precision electroweak parameter $S$ as a function of the hyperscaling coefficient.
We find a general bound on the mass of the lightest spin-1 resonance, by requiring 
that the indirect bounds on the precision parameters be satisfied, that implies that precision electroweak physics
excludes the possibility of a techni-rho meson with mass lighter than several TeV.
\end{abstract}

\maketitle

\tableofcontents

\section{Introduction}

The LHC collaborations published convincing evidence of the discovery of the Higgs particle~\cite{ATLAS,CMS}.
It has mass close to 125 GeV.
  Its couplings are compatible with the predictions
of the minimal version of the Standard Model (SM)~\cite{Higgscouplings}.
And there is no  evidence for any other new particles besides the SM ones.
All of these facts have caused an important shift in the way we think about models of electroweak symmetry breaking (EWSB),
as they exacerbated the general theoretical shortcomings of both the main avenues along which model building
developed for many decades. 
On the one hand, the vast majority of dynamical models of electroweak symmetry breaking (DEWSB), 
in which the underlying dynamics is strongly coupled~\cite{TC,reviewsTC},
 are  incompatible with experimental data, given that they do not yield a light,
 narrow, composite scalar state in the spectrum. On the other hand,
the generic weakly coupled (UV incomplete) models are highly fine-tuned.

The former statement is self-explanatory.
The latter represents a failure in the application of the principles of low-energy effective theory,
that can be interpreted as the statement that only rare, special UV completions
of the weakly coupled models are actually realistic. Hence it is a different formulation of the same
problem as in strongly-coupled models: EWSB as we experience  it in collider physics
must originate from UV physics that is not generic, but rather has unusual properties.
In this sense, the current LHC program is already providing us with interesting (though unclear)  insight about the 
more fundamental theory ruling short-distance physics.
It is now more interesting than ever to focus our attention
on the special and rare classes of 
DEWSB models that are compatible with the data, and try to characterize them better.
In the next round of data that the LHC is going to collect and analyze, some of these models will be tested and possibly excluded
experimentally.

Within DEWSB models, the discovered Higgs particle can be reinterpreted as a composite (pseudo- and techni-) {\it dilaton},
originating from the fact that the underlying dynamics is intrinsically multi-scale and 
approximately scale invariant over some range of energies.
This is an old idea~\cite{dilaton}, which has been proposed in connection with walking technicolor~\cite{dilaton,WTC}
and yields remarkable phenomenological features~\cite{dilatonpheno}.
The characterization of such  models by traditional techniques of quantum field theory  is 
 hard, because of the underlying strong coupling dynamics and because of the non-trivial, multi-scale,
near-conformal dynamics required by phenomenological considerations~\cite{dilaton4}.
It attracted a lot of renewed interest in recent times~\cite{dilatonnew}, 
especially in direct connection with the LHC data~\cite{dilatonandpheno}.
 Lattice studies of models that have non-QCD-like features of relevance for
 DEWSB have in recent years become possible,
and some significant progress has been achieved~\cite{lattice,latticeHiggs}.

The advent of gauge-gravity dualities~\cite{AdSCFT,reviewAdSCFT} offers an independent, complementary framework
 for this exploration: it is possible to study the
weakly coupled dynamics of certain gravity models in extra dimensions and interpret it in terms of special
strongly-coupled four-dimensional gauge theories, by applying a set of rules (dictionary)
that allow to read off correlation functions between operators of the gauge theory.
One branch of the broad program of theoretical exploitation of this tool involves applying it
systematically to the physics of EWSB, in particular in order to classify what type of strongly-coupled field theories exist
that yield phenomenologically viable models, testable at the LHC and future colliders.
There exist at least three broad classes of such models yielding a {\it holographic techni-dilaton} in the spectrum,
 each of which has its own specific advantages and limitations.

 Within the bottom-up approach to holography, already in simple generalizations of the Randall-Sundrum scenario~\cite{RS1} 
supplemented by the Goldberger-Wise mechanism~\cite{GW} one expects a light scalar in the spectrum, related to the 
stabilization of the extra dimension---the radion~\cite{dilaton5D}. The low-energy physics of the radion coming from extra
dimensions coincides (at leading order) with that of the dilaton of the four-dimensional theory.
Many phenomenological models of EWSB of this class exist and have been studied, 
in which the departure  of the background from AdS is 
controllably small~\cite{EP,TCholobottom,Barcelona}.
This requirement is put in by hand within this framework, by arranging for the classical higher-dimensional sigma-model coupled to gravity to yield 
appropriately small effects. This approach has the advantage of being flexible and easy to implement.
But it comes with the disadvantage that the models under consideration cannot be considered as fundamental,
while as effective theories they are  fine-tuned~\cite{RZ}. The importance of these toy models resides in the fact
 that thanks to their simplicity they allow to explore 
what are the open possibilities, and they hence provide a guidance towards building 
more robust, realistic, models.\footnote{There exists also a different variation of the bottom-up approach, in which one builds the 
gravity model in such a way as to reproduce by hand some specific desired features inferred from 
field theory arguments. This is done by choosing appropriately complicated sigma-model potentials
in order to reproduce QCD-like features in the resulting putative dual field theory (see for instance~\cite{Kiritsis}, and references therein).
While interesting for other purposes, these models do not exhibit the presence of a light scalar in the spectrum, and are in
 general not suited for the description of EWSB~\cite{Kiritsis2}.}

In models in which the extra-dimension background is given by solving the equations derived from a complete gravity theory 
(string theory), a comparatively simple and flexible way to discuss phenomenological aspects related to EWSB
consists of introducing a (small) set of extended objects ($Dp$-branes), and studying them 
in probe approximation~\cite{TCholoprobe}, to look for cases in which the classical solution for the embedding leads to
phenomenologically useful results (the spontaneous breaking of chiral symmetry, for example). 
The spectrum of fluctuations of the resulting system contains in some cases
a light scalar, that can be identified as the dilaton. 
This gives rise to useful phenomenological models that are well-defined from the gravity point of view, and 
hence are close to being complete. Yet, they still do not offer a solution to the original problem of providing a fully satisfactory UV-complete
dynamical origin for EWSB and for the light Higgs particle. 
The reason can be better understood by thinking in terms of the putative dual field theory.
Within this class of models, the dilaton is not part of the spectrum of the underlying strong dynamics,
the dual of which is the curved gravity background.
Rather, it originates from an additional 
weakly-coupled sector of the theory (the probes). 
Hence, even if one can identify a light scalar state, it is not clear that such state 
would be protected against quantum corrections due to the further addition 
of the weakly-coupled SM fields. 
This is a serious concern: even in the SM itself, loops of top quarks and other weakly-coupled fields
are those that provide the largest 
quantum corrections that render the Higgs potential fine-tuned.

There is a third class of models, within the top-down approach to gauge/gravity dualities. 
One can look for solutions of the underlying gravity equations 
 exhibiting special features that can be used to improve the phenomenology,
such as  the existence of several dynamical scales~\cite{TCholotop,ENP1,ENP2,EKS, KMPP, EFHMP}. 
In some special cases, the analysis of the spectrum 
of fluctuations of the background metric (having included all mixing and back-reacted all other fields) 
yields a parametrically light scalar state~\cite{ENP1,ENP2,EKS}. This state emerges from the strong dynamics itself, as a 
composite object, the lightness of which descends from the multi-scale dynamics and its non-linearly
 realized (and approximate) scale  invariance. As such the addition of a weakly-coupled sector including the SM fields 
 would not reintroduce a fine-tuning problem.
 
 A technical subtlety, related to this physical point, demands for a clarification.
 On the one hand, it is always possible,  for any gravity dual of any strongly coupled dynamical model, to choose the weak sector to make one of the composite scalar states arbitrarily light~\cite{FFS}, at the price of fine-tuning the 
  boundary action in the gravity dual~\cite{EFHMP}. 
Yet,  calculations such as those in~\cite{ENP2} assume the worst case scenario in 
which divergent localized mass terms are added to the action, and hence
  the finding of a light state is per se a result that is independent of the boundary mass itself.
  Equivalently, this means that  the lightness of the dilaton cannot be spoiled simply by the addition of a weakly-coupled sector,
  and no fine-tuning is present, in spite of the fact that the precise value of the mass does depend on the weakly-coupled
  sector of the theory as well as on the strongly-coupled one.

The technical difficulty in finding models of the three classes outlined above increases dramatically 
 going towards a complete top-down model.
The purpose of this paper is to use a subclass of simple models of the first type described above (bottom-up),
  in order to offer more clear guidance towards how to construct (and what to expect from) models of the third type (top-down).
In particular, we want to address three
specific problems.
\begin{itemize}
\item[1)] Only one class of confining models has been identified within the top-down approach to holography 
that exhibits multi-scale dynamics, walking and a parametrically light composite scalar~\cite{ENP1,ENP2}. 
It is not known what is the general systematics
yielding this result, nor is it known whether other models  exist that  yield these results.
\item[2)] The known models have geometries that are nowhere near AdS, 
but rather exhibit hyperscaling violation (see the body of the paper for an explanation of what this means)
over some intermediate range of the radial direction in the geometry,
and it is not known what this means in field-theory terms.
It is also not  clear how this finding relates to the bottom-up context, 
given that most studies are restricted to near-AdS geometries.
\item[3)] The overwhelming complication of the known models within the top-down approach has so far prevented the construction of a realistic model on its basis, and the study in full of its phenomenological implications.
\end{itemize}

The class of models we study in this paper is characterized by the fact that the gravity background
exhibits hyperscaling violation, at least asymptotically.
 In reference to the three problems just mentioned, we want to understand 
what choices of hyperscaling coefficient $\theta$ yield a dilaton in the spectrum (problem 2),
whether this result can survive the presence of a possible singularity in the geometry,
and whether it is possible to use such backgrounds to construct  semi-realistic models of EWSB (problem 3).
As we will see, the answers we find are quite non-trivial, and yield some guidance 
towards addressing also problem 1,  in the process of finding new larger classes of useful solutions within
the full top-down approach to holography.

Before moving further, the reader should be alerted about a subtlety.
All the five-dimensional Poincar\'e domain-wall background solutions 
for any kind of scalar sigma-model coupled to gravity
that contain an end-of-space (in what is interpreted as the IR regime of the dynamics)
are singular. Such a singularity in the five-dimensional background
 is not necessarily fatal. Many examples exist 
 (the most famous being~\cite{Witten,KS,MN,Butti})
in which the five-dimensional singularity is resolved by lifting to 10 dimensions within a known supergravity theory
of which the five-dimensional sigma-model is a consistent truncation.
Hence, the five-dimensional  IR singularities have little meaning per se,
and in this paper  we are simply not going to concern ourselves with their  nature,
aside from mentioning that all the IR singularities we find are good singularities
when confronted with the criterion proposed by Gubser~\cite{Gubser}.
We will be a bit more careful with the UV singular behavior, for a set of reasons
to be discussed later on, among which the fact that 
gaining control over the UV-asymptotics  is crucial in order to perform the holographic renormalisation procedure~\cite{HR},
which is necessary
to retain a massless photon in the spectrum and hence to produce a semi-realistic toy model of EWSB.

The paper is organized as follows.
Section II contains the basic definitions of the five-dimensional model, and the main classes of solutions we want to investigate.
We discuss a class of solutions that exhibit hyperscaling violation, 
and two classes of solutions in which hyperscaling violation
appears only asymptotically in the UV of the geometry,
while in the IR the geometry ends. We call  the former HSV, and refer to the latter as confining solutions (with abuse of language).
Section III is devoted to the study of the spectrum of scalar fluctuations in the model. We compare the results obtained in the HSV case 
(where a mass gap is  introduced by hand via a physical hard-wall in the geometry), and the confining case.
Section IV shows a special realization of a weakly gauged $U(1)\times U(1)$ theory, higgsed to the diagonal $U(1)$,
and discusses its basic phenomenology, as well as the calculation of the $S$-parameter~\cite{Peskin,Barbieri}.
Section V contains a final discussion about the implications of this study.
We also include three Appendices in which we relegate some technical points.

\section{The model}


We consider a five-dimensional sigma-model coupled to gravity. The field content consists of a $U(1)_L\times U(1)_R$ gauge symmetry 
with gauge bosons $L_M,R_M$ and field strengths $F_{MN}$, as well as one
(neutral)  real scalar field $\chi$. 
We write the bulk action as
\beqs
\label{Eq:S5}
{\cal S}_5&=& \int \di^5 x \sqrt{-g_{5}} 
\left(\frac{{\cal R}_5}{4}\,-\,\frac{1}{2}g^{MN}\partial_M\chi\partial_N\chi \,+\,\Lambda e^{-2\delta \chi}
\,-\,\frac{\kappa\,e^{\omega \chi}}{4}g^{MR}g^{NS}F_{MN}F_{RS}\right)\,,
\eeqs
so that the potential is $V=-\Lambda e^{-2\delta \chi}$, with $\Lambda$ a constant.
We use capital Roman indexes $M\,=\,0\,,\,1\,,\,2\,,\,3\,,\,5$ in five dimensions and Greek indexes
$\mu\,=\,0\,,\,1\,,\,2\,,\,3$ in four.
The model depends on the real parameters $\Lambda$, $\omega$, $\kappa$ and $\delta$,
the meaning of which will be explained in due time, together with the restrictions we will be forced to impose on the freedom
in their choices.

We use the Poincar\'e-invariant domain-wall ansatz for the metric
\beqs
\di s_5^2 &=& e^{2A(r)} \di x^2_{1,3}\,+\,\di r^2\,,
\eeqs
where $\di x^2_{1,3}$ is the 4-dimensional Minkowski metric with signature $\{-\,,\,+\,,\,+\,,\,+\}$.
Furthermore, we assume that  the scalar $\chi$ admits a non-trivial $r$-dependent background configuration $\chi(r)$,
but we do not allow the gauge bosons to have such a non-trivial classical configuration.

We perform the change of variable  $\partial_r=e^{-\delta \chi} \partial_{\r}$, so that
the metric is 
\beqs
\di s^2_5 &=& e^{2A(\r)} \di x_{1,3}^2\,+\,e^{2 \delta \chi(\r)} \di \r^2\,.
\eeqs
The equations of motion in these coordinates are
\beqs
0&=&4  A'(\r) \chi '(\r)-\delta  \left(2 \Lambda + \chi '(\r)^2\right)+\chi ''(\r)
\,,\\
   0&=&3 A''(\r)-3 \delta  A'(\r) \chi '(\r)+6 A'(\r)^2-2 \Lambda +\chi '(\r)^2
   \,,\\
   0&=&6 A'(\r)^2-2 \Lambda -\chi '(\r)^2
   \,,
\eeqs
where the $'$ represents derivatives with respect to $\rho$. We interpret the resulting model as the dual of a four-dimensional field theory. The radial coordinate $r$ corresponds to renormalisation scale, such that large-$r$ corresponds to the UV and small-$r$
to the IR. We assume that the gauge/gravity dictionary can be applied to the resulting system,
and use it to calculate physical observables in the putative four-dimensional theory.

These coupled non-linear differential equations admit several different classes of solutions.
In this paper we focus on two special classes. As we will discuss, we will be forced to make some special choices for
the model parameters in order for the solutions to be well defined and to admit a sensible physical interpretation.

For reasons that will become clear later, we  write
\beqs
\delta&\equiv&\sqrt{\frac{-2\theta}{3(3-\theta)}}\,,
\eeqs
where, because both the metric and the action depend only on the combination
$\delta \chi$, we  chose $\delta>0$ without loss of generality, while we have to be careful about the
sign of $\chi$.
We also choose to fix the constant $\Lambda$ to be given by 
\beqs
\Lambda&\equiv&\frac{\theta-3}{4(\theta-4)}\,,
\eeqs
for reasons that will become clear later.
This specific form amounts to setting the units in terms of an overall scale, hidden within the definition of $\Lambda$.
As we will see, we are only interested in ratios of scales, and hence there is no loss of generality.

\subsection{HSV solutions}

The simplest class of solutions to the equations is of the form
\beqs
\chi&=&
\chi_0\,+\,\gamma\,\r\,,\\
A&=&A_0\,+\,\alpha\,\r\,,
\eeqs
with $\chi_0$ and $A_0$ integration constants.
By replacing into the equations of motion, the resulting algebraic system is solved by
\beqs
\alpha&=&
s\,\sqrt{\frac{8\Lambda}{3(8-3\delta^2)}}\,=\,
s \sqrt{\frac{(3-\theta)^2}{9(4-\theta)^2}}\,,\\
\gamma&=&
s\sqrt{\frac{6\Lambda\delta^2}{8-3\delta^2}}\,=\,
s \sqrt{\frac{\theta(\theta-3)}{6(\theta-4)^2}}\,,
\eeqs
where $s=\pm 1$.
From these expressions we can see that the solution is real for the choice of $\Lambda$ we adopted.
The special choice $\theta=0$ corresponds to $\delta=0$, $\theta=3$ corresponds to $\Lambda=0$,
and  $\theta=4$ corresponds to the limiting case $\delta^2=8/3$.

The choice of positive $\alpha$ and $\gamma$ is such that by defining the quantity~\cite{FGPW}
\beqs
C&=&\frac{1}{(\partial_rA)^3}\,=\,\frac{1}{(e^{-\delta \chi}\partial_\r A)^3}\,,
\eeqs
which is related to the central change of the (putative) dual field theory,
we would ensure that $C>0$ and $C^{\prime}>0$. This would be in agreement with the expectations from the $C$-theorems
according to which the central change is positive and monotonically non-increasing along RG flows towards the IR.
We will see that we cannot do so: while for $\theta<0$ and $\theta>4$ we can indeed choose $s=+1$,
 for $3<\theta<4$ we will be forced to choose $s=-1$ in order for the HSV solutions to be a sensible approximation 
 (for large $\r$) of the solutions in the other classes we consider. 
 As  a consequence, while we always work with backgrounds for which $C^{\prime}>0$, 
 in the range $3<\theta<4$ we have $C<0$. We do not comment further on this, but the reader should be alerted that results
 for $3<\theta<4$ should be taken carefully.

We can write the metric in  conformal form, by means of the change of variable
\beqs
\di \r &=& \frac{1}{\gamma \delta - \alpha} \,\frac{\di z}{z}\,,
\eeqs
and the metric, after a rescaling of the Minkowski coordinates, becomes
\beqs
\di s_5^2 &\propto&
z^{-2+\frac{2}{3}\theta}\,\left(\frac{}{}\di x_{1,3}^2\,+\,\di z^2\right)\,.
\eeqs
Under the transformation
\beqs
x^{\mu}, z&\rightarrow \lambda x^{\mu},\lambda z\,,
\eeqs
one finds that
\beqs
\di s^2_5&\rightarrow & \lambda^{\frac{2\theta}{3}}\di s_5^2\,,
\eeqs
which is what is meant by hyperscaling violation (or sometimes referred to in the literature  as scale covariance).
We see that we are adopting the same conventions as in~\cite{Dong:2012se} for the hyperscaling coefficient $\theta$,
hence explaining to the reader the way in which we redefined the parameter $\delta$.
This type of metrics can be obtained for example from compactifying higher dimensional theories, 
as done in~\cite{Perlmutter,EFHMP}.
The expression for $\delta$ is well-defined (real) only for $\theta\leq 0$ or $\theta> 3$,
 in agreement with the analysis 
of the null energy condition in~\cite{Dong:2012se}, from which it emerges that in the range $0<\theta<3$ there is an instability.
Indeed, in order to adopt values of $0<\theta<3$, the potential in our model would not be real.

\subsection{Confining Solutions}

There exist more general classes of solutions, in which HSV behavior is recovered only 
for large $\r$, while the geometry ends non-trivially at some finite value of $\r$.
We write these solutions in the following way:
\beqs
\label{eq:confiningbackgroundA}
A(\r)&=&
A_0\,+\,\frac{(\theta -3) \log (\sinh (\r))}{3 (\theta -4)}+\frac{\theta\sqrt{(\theta -3) /\theta } \log \left(\tanh
   \left(\frac{\r}{2}\right)\right)}{6 (\theta -4)}
   \,,\\
\label{eq:confiningbackgroundchi}
\chi(\r)&=&\chi_0\,+\,
\frac{\sqrt{\frac{3}{2}} \log \left(\sinh \left(\frac{\r}{2}\right)\right)}{2-\sqrt{\frac{\theta }{\theta
   -3}}}-\frac{\sqrt{\frac{3}{2}} \log \left(\cosh \left(\frac{\r}{2}\right)\right)}{\sqrt{\frac{\theta }{\theta
   -3}}+2}\,,
\eeqs
where $A_0$ and $\chi_0$ are integration constants,
and we fixed another integration constant by the requirement that the space ends at $\r\rightarrow 0$.
The equations cannot be extended to the range $0<\theta<3$ (which we know is an unphysical region),
because $\chi$ and $A$ must be real.

We call these solutions {\it confining}, with abuse of language: what might confine is the dual field theory,
and even that is in general not true (but there are examples, such as those discussed in~\cite{EFHMP,F4QCD4}). 
The rigorous statement would be  that the gravity solutions have an end-of-space at $\r=0$,
at which the five-dimensional geometry becomes singular, and that the putative dual theory becomes trivial
below the four-dimensional scale corresponding to $\r=0$. This can be the effect of confinement, but more in general just 
implies the presence of an IR cutoff. As anticipated, for this class of solutions the central charge $C$ is monotonically 
increasing towards the UV, but negative for $3<\theta<4$.

There is  another solution,
 in which some of the terms in $\chi$ and $A$ 
have opposite sign (it corresponds to switching $\sinh(\r/2)\leftrightarrow \cosh(\r/2)$ ).  
In the range $3<\theta<4$, the alternative solution has non-monotonic $A(\r)$, and as a consequence $C$ diverges at some finite value of $\r$, signaling the fact that  this solution is unphysical. The alternative solution is:
\beqs
A(\r)&=&
A_0+\frac{(\theta -3) \log (\sinh (\r))}{3 (\theta -4)}-\frac{\theta  \sqrt{(\theta -{3})/{\theta }} \log \left(\tanh
   \left(\frac{\r}{2}\right)\right)}{6 (\theta -4)}\,,\\
   \chi(\r)&=&
\chi _0   -\frac{\sqrt{\frac{3}{2}} \log \left(\sinh \left(\frac{\r}{2}\right)\right)}{\sqrt{\frac{\theta }{\theta
   -3}}+2}+\frac{\sqrt{\frac{3}{2}} \log \left(\cosh \left(\frac{\r}{2}\right)\right)}{2-\sqrt{\frac{\theta }{\theta
   -3}}}\,.
\eeqs
We explicitly checked that both classes of solutions satisfy the criterion in~\cite{Gubser},
 namely the potential $V$ of the five-dimensional sigma-model 
is bounded from above when approaching the IR singularity. We also found that the potential for the alternative solution is always larger than for the confining solution.

We also computed the (free) energy of the putative dual theory, by replacing the solution into the action.
We find that the contribution of the bulk action is
\beqs
E_5(\r)&=&
-\frac{2^{-\frac{2 (\theta -6)}{3 (\theta -4)}} (\theta -3) \sinh (\r)}{3 (\theta -4)}\,,
\eeqs
for both the confining solutions, both for $\theta<0$ and $\theta>4$.
Because we will have to regulate the system, we need to add the boundary terms,
which consist of the Gibbons-Hawking term and a localized cosmological constant 
(we follow the notation in~\cite{EP}):
\beqs
S &=& \int \di^5x \left[\delta (r-r_I)\sqrt{-\tilde{g}}\left(-\frac{1}{2} K-\lambda_I\right)
-\delta (r-r_U) \sqrt{-\tilde{g}}\left(-\frac{1}{2}K-\lambda_U\right)\right]\,,
\eeqs
where 
\beqs
\lambda_i&=&-\frac{3}{2}A'(r_i)\,.
\eeqs
The result is that at the IR boundary ($\r_I$) and UV boundary ($\r_U$) the contributions to the energy are, respectively,
given by
\beqs
E_{IR}&=&\frac{2^{\frac{\theta }{3 (\theta -4)}-2} \left(-\sigma \theta\sqrt{(\theta -3) /\theta }-2 (\theta
   -3) \cosh (\r_I)\right)}{3 (\theta -4)}\,,\\
 E_{UV}&=&-\frac{2^{\frac{\theta }{3 (\theta -4)}-2} \left(-\sigma\theta\sqrt{(\theta -3) /\theta }-2 (\theta
   -3) \cosh (\r_U)\right)}{3 (\theta -4)}\,,
\eeqs
where $\sigma=+1$ for the confining solution and $\sigma=-1$ for the alternative solution.
But notice that the difference between the two solutions appears only in terms that are independent of 
$\r_i$, and hence cancel exactly in the total energy.
We conclude that 
\beqs
E_T&=&E_{IR}+E_{UV}+\int_{\r_I}^{\r_U} \di \r E_5(\r) = 0
\eeqs
exactly agrees for the two solutions.
Nevertheless, we anticipate here that 
 that the study of the spectrum of scalar fluctuations reveals the presence of a tachyon for the alternative solution, which suggests that this second class of solutions should be disregarded.
 We leave it to a future study to enquire on a more general dynamical explanation for this fact.

\section{Spectrum of scalar bound states}

We want to compute the mass spectrum 
of scalar bound states of the dual four-dimensional gauge theory, by studying the fluctuations of the $\sigma$-model 
 coupled to gravity in five dimensions.
The fluctuations of the  scalars  and of the metric couple in a quite non-trivial way.
Furthermore, because of  diffeomorphism invariance of the gravity theory, many combinations of the
original scalar and gravity fluctuations are spurious, and must be removed.
And also, because some of the background 
functions diverge in the UV and in the IR, it is necessary to introduce two regulators $0<r_I \ll r_U$,
 solve for fixed $r_I$ and $r_U$, and then take the limit $r_U\rightarrow +\infty$ and
$r_I\rightarrow 0$.

In the light of all these technical difficulties, it 
is convenient to use the gauge-invariant formalism  developed in~\cite{BHM}, 
and extended in~\cite{E} to the general case in which the $\sigma$-model does not admit a 
superpotential. We apply at $r_I$ and $r_U$ the boundary conditions discussed in~\cite{EP},
 taking the special limit in which divergent boundary masses are added for the $\sigma$-model scalars. 
 After the regulators are removed, this effectively enforces the correct boundary conditions 
 (regularity in the IR and normalizability in the UV) dictated by holography, and 
 ensures that if we find a light state this is going to be a generic result, not an artifact of a special choice of boundary
 conditions.
 
The main result of the analysis in~\cite{EP} is that, given a $\sigma$-model with $n$ scalars and 
action of the form we have, the physical fluctuations are given by gauge-invariant scalar functions denoted by $\mathfrak{a}^a$ and satisfying the bulk equations 
\beqs
\label{Eq:diffeq}
	0&=&\Big[ {\cal D}_r^2 + 4 A' {\cal D}_r + e^{-2A} \Box \Big] \mathfrak{a}^a \\ \nonumber
	&& - \Big[ V^a_{\ |c} - \mathcal{R}^a_{\ bcd} \bar \Phi'^b \bar\Phi'^d + \frac{4 (\bar \Phi'^a V_c + V^a \bar \Phi'_c )}{3 A'} + \frac{16 V \bar \Phi'^a \bar \Phi'_c}{9 A'^2} \Big] \mathfrak{a}^c ,
\eeqs
while the boundary conditions are given by
\beqs
\label{Eq:BCb}
	&\left[  e^{2A} \Box^{-1}  \frac{2  \bar \Phi'^c }{3 A'} \right] 
	\left( \bar \Phi'_b{\cal D}_r -\frac{4 V \bar \Phi'_b}{3 A'} - V_b \right)\mathfrak a^b \Big|_{r_i} =  
	-\mathfrak a^c \Big|_{r_i}.
\eeqs
In these expressions, $^{\prime}$ indicates derivatives in respect to $r$. $A$, $\bar{\Phi}^a$ and their derivatives refer to the 
background fields, $V_b=\partial V/\partial \Phi^b$ is a field derivative of the potential, $\mathcal{R}^a_{\ bcd}$ is the Riemann tensor 
associated with the $\sigma$-model metric, ${\cal D}_r$ is the background covariant derivative, and $V^a_{\ |c}$ is the ($\sigma$-model) 
second-order covariant derivative. Details and explanations about the notation can be found in~\cite{BHM, EP}.

In the present case, the algebra significantly simplifies, thanks to the fact that we have only one scalar,
and the sigma-model metric is trivial. Remember that we need to change variable according to $\partial_r=e^{-\delta\,\chi}\partial_{\rho}$.

\subsection{HSV solutions}

In the case of HSV solutions (and setting $\chi_0=0$) we find that the 
bulk equation reduces to 
\beqs
\mathfrak{a}^{\prime\prime}\,+\,\mathfrak{a}^{\prime}\,+\,e^{\frac{2\r}{\theta-4}}q^2 \mathfrak{a}&=&0\,,
\eeqs
and the boundary conditions to
\beqs
\left.\frac{}{}\mathfrak{a}^{\prime}+\frac{3(\theta-4)}{\theta}e^{\frac{2\r}{\theta-4}}q^2 \mathfrak{a}\right|_{\r_i}&=&0\,.
\eeqs

\begin{figure}[h,t]
\begin{center}
\begin{picture}(260,160)
\put(10,7){\includegraphics[height=5cm]{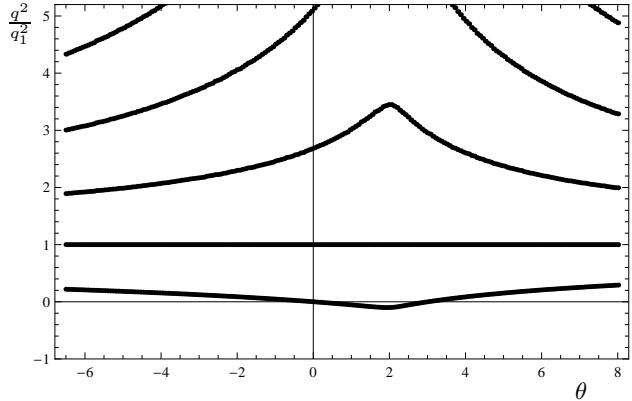}}
\put(0,140){$\frac{q^2}{q^2_1}$}
\put(215,0){$\theta$}
\end{picture}
\caption{Mass $q^2/q^2_1$ of the scalar glueballs as a function of $\theta$
in the hyperscaling case.
In the numerics, we used $\r_U=20$ and $\r_I=0.01$.}
\label{Fig:scalarstheta}
\end{center}
\end{figure}

We performed a numerical study, the results of which are shown in Fig.~\ref{Fig:scalarstheta}. We computed the spectrum by 
extending the bulk equations and boundary conditions also in the unphysical range $0<\theta<3$, 
where (unsurprisingly) we find a tachyon.
Here and in the following, we show the spectrum of mass eigenstates normalized so that the  second mass state
(denoted $q_1$) is set to $q_1=1$. The reason for doing so is that the main thing we are interested in understanding
is whether there is a sense in which the lightest state in the theory is parametrically lighter than the typical mass of the higher excitations.

Our main result is that we find two regions for which a parametrically light state appears in the spectrum of the HSV case.
One is for backgrounds that are close to AdS$_5$, and have $-1\ll\theta\leq 0$.
The other is for the case in which $3\leq\theta\ll 4$, in which case the hyperscaling coefficient $\theta$ is close to the one of flat space.

\subsection{Confining solutions}

\begin{figure}[h,t]
\begin{center}
\begin{picture}(480,120)
\put(325,8){\includegraphics[height=3.2cm]{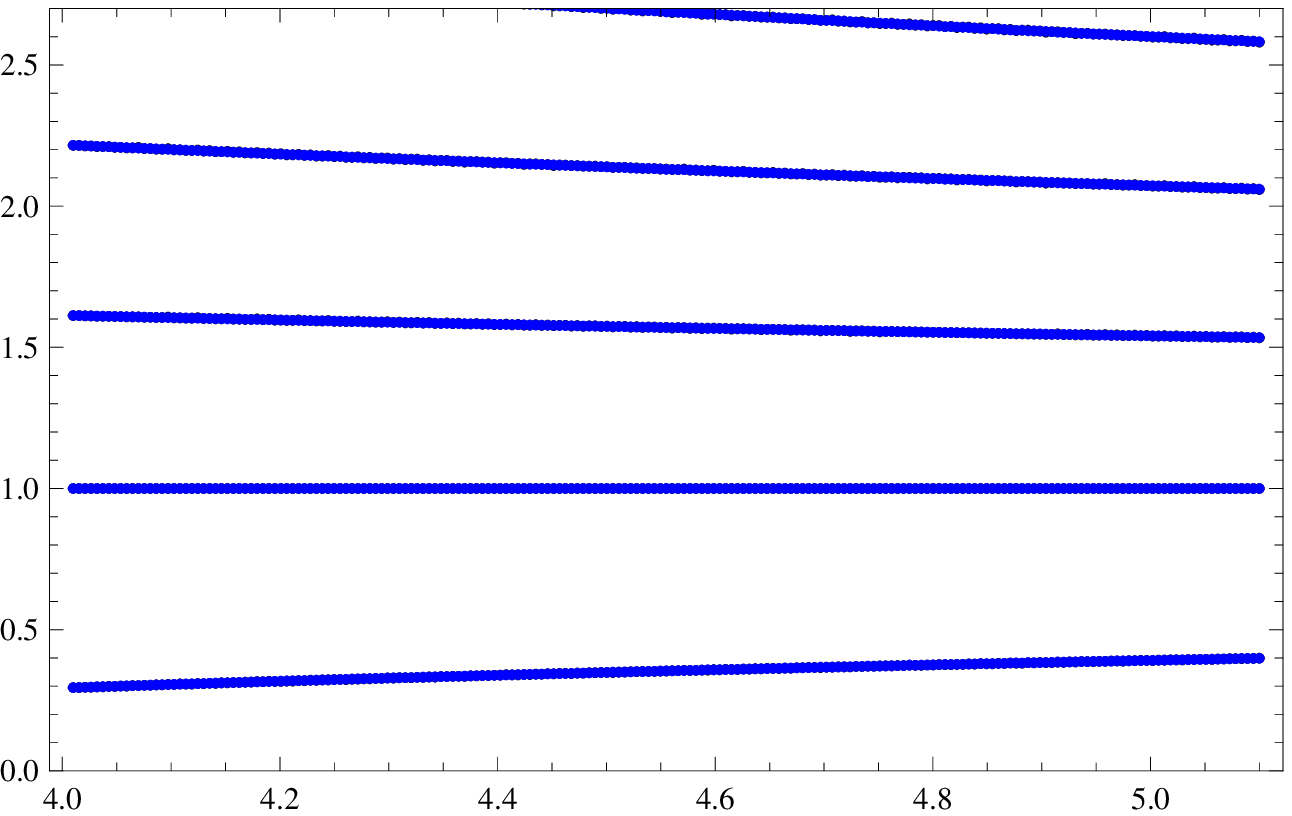}}
\put(5,8){\includegraphics[height=3.2cm]{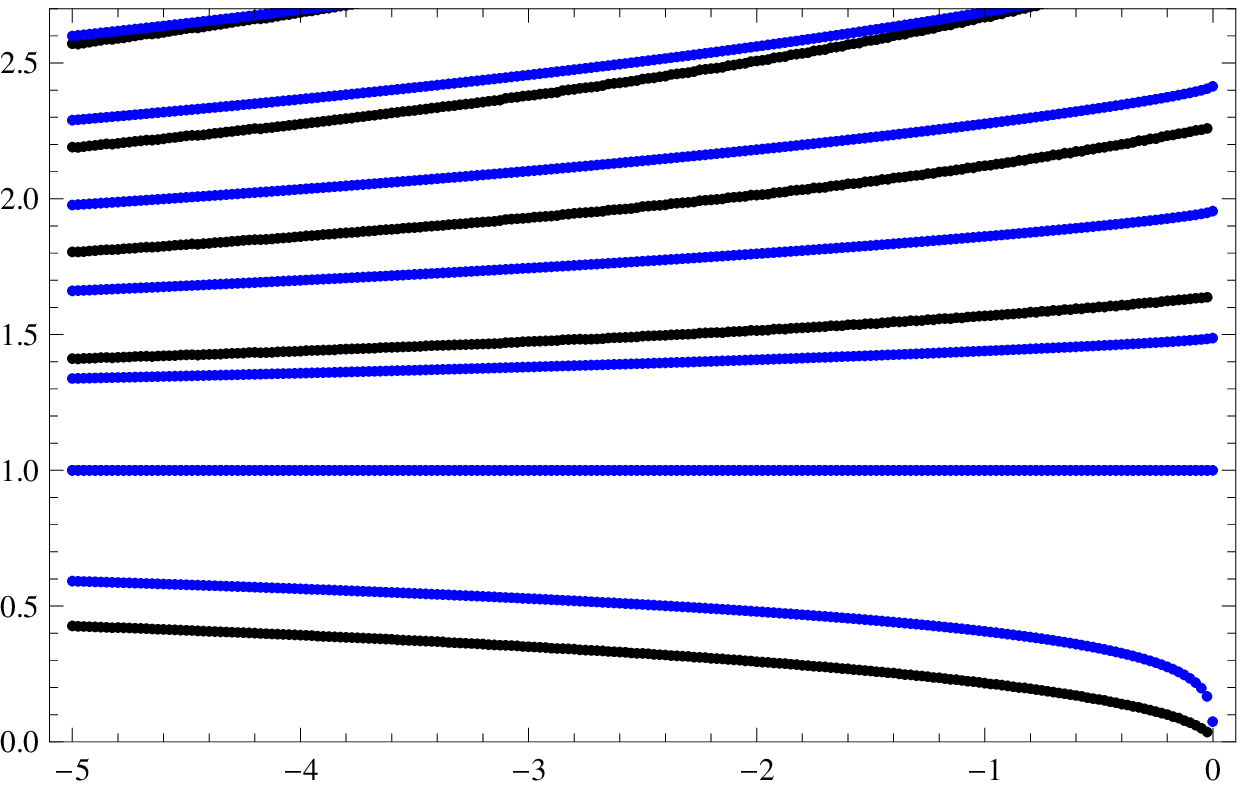}}
\put(165,8){\includegraphics[height=3.2cm]{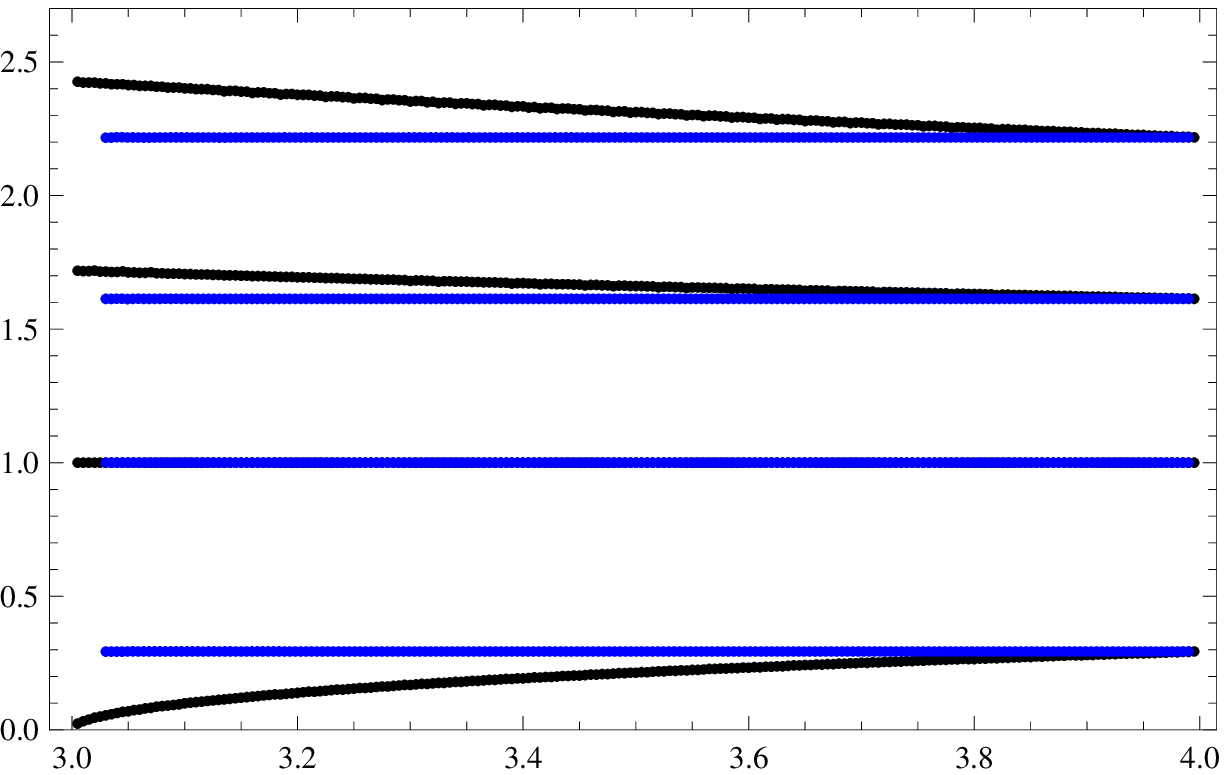}}
\put(0,103){$q/q_1$}
\put(120,0){$\theta$}
\put(160,103){${q/q_1}$}
\put(280,0){$\theta$}
\put(320,103){$q/q_1$}
\put(440,0){$\theta$}
\end{picture}
\caption{$q/q_1$ for the scalar fluctuations as a function of $\theta$
in the confining case (blue) compared with the HSV case (black).}
\label{Fig:scalarsthetaconf}
\end{center}
\end{figure}

We repeat the calculation of the spectra for the confining solutions, which yield the results in Fig.~\ref{Fig:scalarsthetaconf}.
Shown are the three ranges $\theta<0$, $3<\theta<4$ and $\theta>4$, explicitly comparing the results of the confining 
backgrounds to the HSV ones. In Appendix A, we write explicitly the bulk equations of motion as well as boundary conditions for the fluctuations. The UV (IR) cutoff is chosen to be sufficiently high (close to the end-of-space), so as to ensure that the spectrum obtained is not significantly affected by unphysical cutoff effects. In Appendix B we show a sample of the study of the spectrum as it depends on the UV and IR cutoffs.

We find that for $\theta<0$ the mass of the lightest state is larger for the confining solutions 
compared to the HSV case, and yet it stays true that for $\theta\simeq 0$ we find a parametrically light state in the spectrum.
By contrast, there is very little visible effect in the spectrum for $\theta>4$, that closely resembles the HSV case.
The case $3<\theta<4$ is somewhat surprising.
While in the HSV case there is a parametrically light scalar state for $\theta\simeq 3$, this is not true once the effects
of the end-of-space in the geometry is taken into account with the confining solutions.
Furthermore, it appears that in the range $3<\theta<4$ the mass spectrum of scalar states is $\theta$-independent, and
agrees with the HSV case with $\theta=4$.

The case $\theta=3$ is somewhat pathological: in this case $A$ and $\chi$  become constant,
and the space is flat. In this case the only scale in the system is actually $\r_U-\r_I$, the size of the compactified fifth
 dimension, and hence there is a pathology in the limit  $\r_U\rightarrow +\infty$.
 As a result, numerically we were not able to come very close to $\theta=3$, and hence to show that 
 eventually there is a massless state, as in the HSV backgrounds.

\begin{figure}[h,t]
\begin{center}
\begin{picture}(320,120)
\put(5,8){\includegraphics[height=3.2cm]{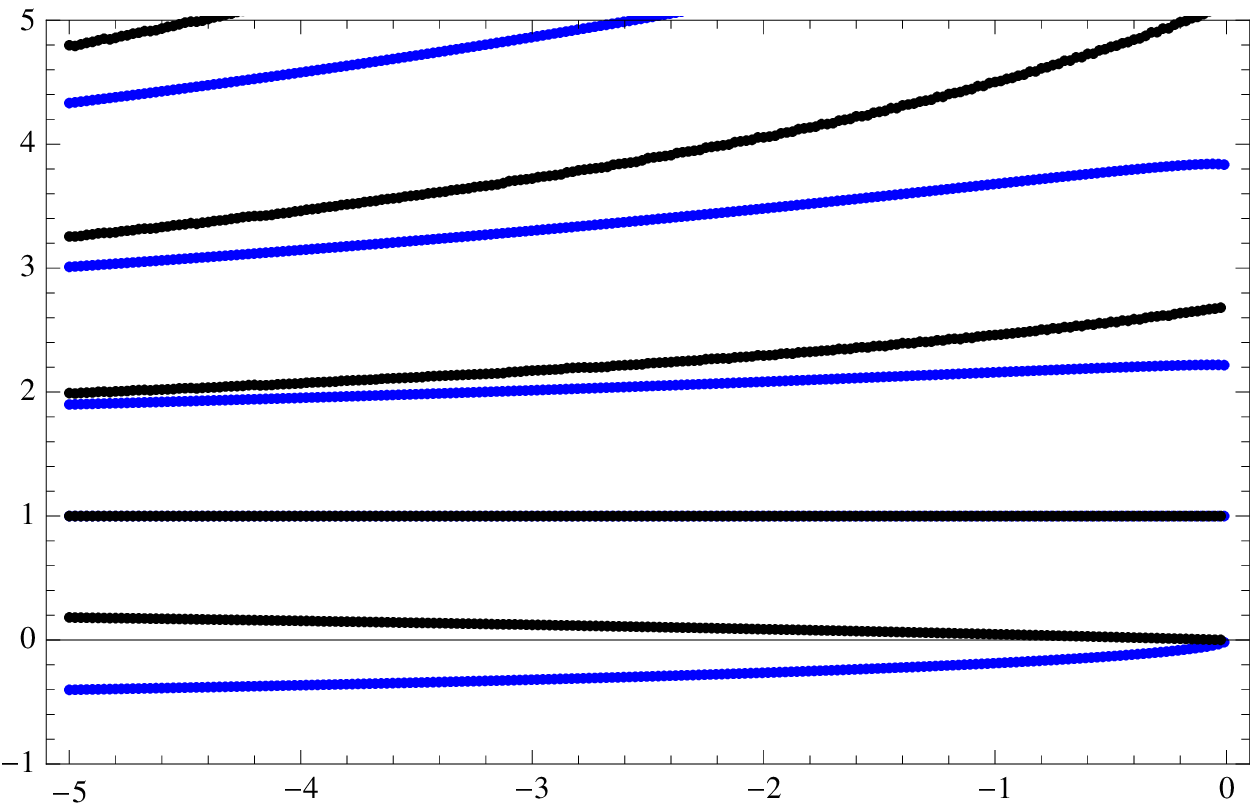}}
\put(165,8){\includegraphics[height=3.2cm]{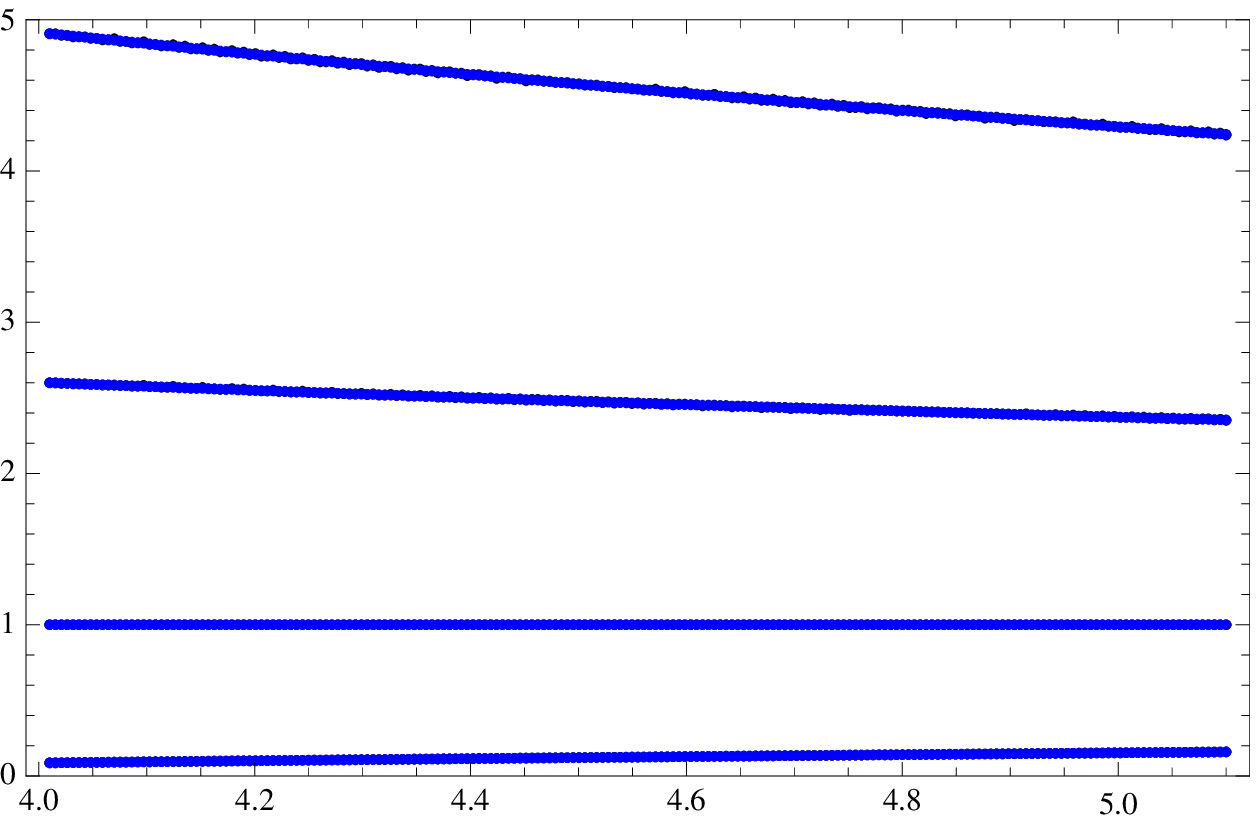}}
\put(0,103){$q^2/q_1^2$}
\put(120,0){$\theta$}
\put(160,103){${q^2/q_1^2}$}
\put(280,0){$\theta$}
\end{picture}
\caption{$q^2/q_1^2$ for the scalar fluctuations as a function of $\theta$
in the confining case (blue) compared with the HSV case (black) for the alternative confining backgrounds.}
\label{Fig:scalarsthetaAlt}
\end{center}
\end{figure}

\begin{figure}[h,t]
\begin{center}
\begin{picture}(260,160)
\put(10,7){\includegraphics[height=5cm]{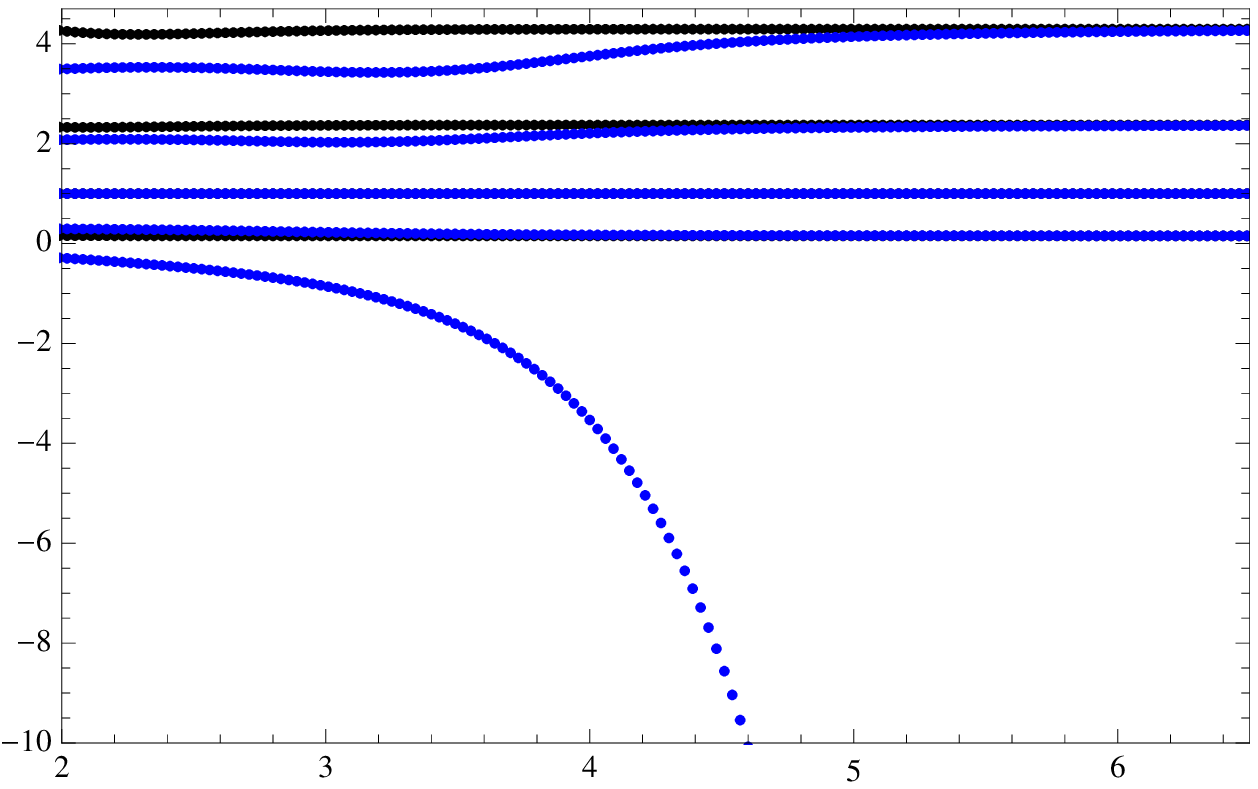}}
\put(0,140){$\frac{q^2}{q^2_1}$}
\put(215,0){$\r_U$}
\end{picture}
\caption{$q^2/q_1^2$ for the scalar fluctuations as a function of $\r_U$ for $\theta=5$ in the confining case (blue) compared with the HSV case (black) for the alternative confining backgrounds.}
\label{Fig:alttachyonrUV}
\end{center}
\end{figure}

We also repeat the exercise for the case of the alternative confining solutions, 
with the only major difference that we are forced to restrict attention to $\theta<0$ and $\theta>4$.
The results are shown in Fig.~\ref{Fig:scalarsthetaAlt}. Notice that we plot $q^2/q_1^2$ because in this case we find a tachyon. It is directly visible for $\theta<0$, while for $\theta>4$ its mass diverges for $\r_U\rightarrow +\infty$,
as can be seen in Fig.~\ref{Fig:alttachyonrUV}. This illustrates that in order to assess whether a given background is made unhealthy due to the presence of tachyon(s) requires a careful analysis of the dependence of the spectrum on the IR/UV regulators. Note also that this provides an example of a situation in which
the (mild) requirements of the criterion in~\cite{Gubser} are satisfied, and yet the background 
must be discarded.

\section{Electroweak symmetry and symmetry breaking and $S$ parameter}

In this section we construct and study a simplified model of EWSB, in which we focus on the
physics of the neutral gauge bosons, and in which we assume that symmetry breaking be induced
by effects taking place on the IR boundary at $\r = \r_I \rightarrow 0$. We ignore the model-building complications related to the charged gauge bosons because the constraints on the $S$ parameter are the most difficult to evade. We start this section by summarizing some known results, in such a way as to fix the notation. We then apply this to the models introduced earlier on in the paper.

One of the most important bounds on EWSB comes from the $S$ parameter~\cite{Peskin}. We follow the conventions of~\cite{Barbieri},
up to a trivial rescaling of the gauge bosons,
by writing the effective Lagrangian density for the  gauge bosons of $U(1)_L\times U(1)_R$ in  four dimensions as
\beqs
{\cal L}&=&-\frac{1}{2}P^{\mu\nu}{A}^i_{\mu}(q){\pi}_{ij}{A}^{j}_{\nu}(-q)\,+\frac{g^i}{2}J^{i\,\mu}(q){A}^i_{\mu}(-q)
\,+\frac{g^i}{2}J^{i\,\mu}(-q){A}^i_{\mu}(q)\,,
\eeqs
where the index $i=L,R$ labels the gauge couplings $g^i=(g,g^{\prime})$, sources $J^i_{\mu}=(J_{\mu}^L,J_{\mu}^R)$  and gauge bosons $A^i_{\mu}=(L_{\mu},R_{\mu})$
of $U(1)_L\times U(1)_R$, and where $P^{\mu\nu}=-\eta^{\mu\nu}-q^{\mu}q^{\nu}/q^2$
and $q^2\equiv -q^{\mu}q^{\nu}\eta_{\mu\nu}$.
The functions ${\pi}_{ij}(q^2)$ contain all the information about the dynamics, and we 
compute them starting from the five-dimensional theory.
We can expand for small $q^2$ as in
\beqs
{\pi}_{ij}(q^2)&=&{\pi}_{ij}(0)\,+q^2\,{\pi}^{\prime}_{ij}(0)\,+\frac{1}{2}(q^2)^2\,{\pi}^{\prime\prime}_{ij}(0)+\cdots\,.
\eeqs

We adopt conventions such that our results, in the basis $(A_{\mu},V_{\mu})$, take the form:
\beqs
\hat{\pi}&\simeq&\left(\begin{array}{cc}
\pi_A(0)+\pi_A^{\prime}(0)q^2+{\cal O}(q^4)& 0\cr
 0 & \frac{1}{e^2} q^2+{\cal O}(q^4) \cr
 \end{array}\right) \,,\label{Eq:pinormalization}
\eeqs
where by definition
\beqs
e^2&\equiv&g^2\sin^2\theta_W\,\equiv\,\frac{g^2g^{\prime\,2}}{g^2+g^{\prime\,2}}\,.
\eeqs
With these conventions, we find that 
\beqs
\hat{S}  &\simeq&\cos^2\theta_W\left(1-g^2\sin^2\theta_W \pi_A^{\prime}(0)\frac{}{}\right)\,,
\label{Eq:Suseful}
\eeqs
where the derivation is relegated to Appendix C, and where we made the implicit assumption that
all the precision parameters, including $\hat{S}$ itself, as well as all the higher-order 
corrections in terms $O(q^4)$ and above, be small. Within these approximations,
one also sees that $M_Z^2\simeq - \pi_A(0)/\pi_A^{\prime}(0)$.
The experimental bounds we use are from~\cite{Barbieri}, for the light Higgs case, and imply that
$\hat{S}<0.0013\,(0.003)$ at the $1\sigma\,(3\sigma)$ c.l.

\subsection{Five-dimensional analysis}

The general action for the gauge bosons in 5 dimensions, 
with which we model the neutral gauge bosons ($Z$, photon) of the Standard Model and their excitations,
contains $U(1)_L\times U(1)_R$ gauge fields $L_M$ and $R_M$ and reads
\beqs
\label{Eq:Seff4}
{\cal S}_{V,A}&=&-\frac{1}{4}\int\di^4 x\int\di \r \left[\frac{}{}\left(a(\r)-b(\r)D\,\delta(\r-\r_U)\right)\eta^{\mu\nu}
\eta^{\sigma\pi}\left(L_{\mu\sigma}L_{\nu\pi}\,+\,R_{\mu\sigma}R_{\nu\pi}\right)\frac{}{}\right.\\
&&\left.\frac{}{}+\,2b(\r)\eta^{\mu\nu}\left(L_{\r\mu}L_{\r\nu}\,+\,R_{\r\mu}R_{\r\nu}\right)\,
-2b(\r)\Omega^2\delta(\r-\r_I)\eta^{\mu\nu}A_{\mu}A_{\nu}\right]\,,\nonumber
\eeqs
where $L_{MN}$ and $R_{MN}$ are the five-dimensional field-strength tensors of $L_M$ and $R_M$.
Notice that the $\Omega^2$ term is a localized symmetry breaking term.
Because the symmetry breaking mechanism is localized, we can work in the gauge $V_{\rho}=0=A_{\rho}$.
The UV-localized kinetic term controlled by $D$ is responsible for the weak gauging of the symmetry,
as we will see shortly. The fact that the localized kinetic term, as well as the bulk action,
are identical for both $A$ and $V$ is a choice we make so that all the isospin breaking effects come from the IR boundary conditions.

After Fourier transforming, and writing $A^{\mu}(q,\rho)=A^{\mu}(q)v(q,\r)$, 
the bulk equations for the gauge bosons are
\beqs
\partial_{\r}\left(\frac{}{}b(\r)\partial_{\r}v(q,\r)\right)&=&-q^2a(\r)v(q,\r)\,.
\eeqs
They can be rewritten in terms of two functions
$\gamma_{V,A}(q,\r)=\partial_{\r}v_{V,A}(q,\r)/v_{V,A}(q,\r)$ as
\beqs
\partial_{\r}\left(\frac{}{}b(\r)\gamma_{V,A}(q,\r)\right)+b(\r)\gamma_{V,A}(q,\r)^2+q^2\,a(\r)&=&0\,,
\eeqs
with the boundary conditions in the IR
\beqs
\gamma_V(q,\r_I)&=&0\,,\\
\gamma_A(q,\r_I)&=&\Omega^2\,,
\eeqs
for the vector (V) and axial-vector (A) bosons, respectively.

By replacing the solutions into ${\cal S}_{V,A}$, one finds that only  UV-localized boundary terms survives from which we read off the 
(transverse) vacuum polarizations $\hat{\pi}$ of the four-dimensional theory defined earlier on:
\beqs
\hat{\pi}_{V,A}&=&-b(\r_U)\left[\frac{}{}D\,q^2 +\gamma_{V,A}(q,\r_U)\right]\,.
\eeqs
We can expand the functions $\gamma_{V,A}$ for small $q^2$.
Dropping the subscripts $_{V,A}$ we write
\beqs
\gamma(q,\r)&=&\gamma_{0}(\r)+q^2\gamma_{1}(\r)\,+\frac{1}{2}(q^2)^2\gamma_{2}(\r)\,+\cdots\,,
\eeqs
and by replacing in the bulk equations we replace the original equation with a sequence of coupled equations for the functions $\gamma_i(\r)$:
\beqs
\partial_{\r}\left(b\,\gamma_0\right)\,+\,b\,\gamma_0^2 &=&0\,,\\
\partial_{\r}\left(b\,\gamma_1\right)\,+\,2b\,\gamma_0\gamma_1\,+\,a &=&0\,,\\
\partial_{\r}\left(b\,\gamma_n\right)\,+\,b\,\sum_{k=0}^n\left(\begin{array}{c}n\cr k\end{array}\right)\gamma_{n-k}\gamma_k &=&0\,,
\eeqs
where the last expression is valid for $n>1$.
These equations must satisfy the boundary conditions
\beqs
\gamma_{0\,,\,V}(\r_I)&=&0
\eeqs
for the vectors, while the axial vectors have
\beqs
\gamma_{0\,,\,A}(\r_I)&=&\Omega^2\,,
\eeqs
and in all cases with $i>0$
\beqs
\gamma_i(\r_I)&=&0\,.
\eeqs

For the vector bosons, these expressions can all be integrated, at least formally, and
the IR boundary condition is solved provided $\gamma_0=0$ identically.
This yields
\beqs
\gamma_{0\,,\,V}(\r)&=&0\,,\\
\gamma_{1\,,\,V}(\r)&=&-\frac{1}{b(\r)}\int_{\r_I}^{\r}\,\di y\,a(y)\,,\\
\gamma_{n\,,\,V}(\r)&=&-\frac{1}{b(\r)}\sum_{k=1}^{n-1}\left(\begin{array}{c}n\cr k\end{array}\right)\int_{\r_I}^{\r}\,\di y\,b(y)\gamma_{n-k}(y)\gamma_k(y)\,. 
\eeqs

For the axial-vector case, because of the different boundary condition in the IR, the resulting equations are 
 more complicated. Most important in the following is the $q^2$-independent term
\beqs
\gamma_{0\,,\,A}(\r)&=&\frac{b(\r_I)\Omega^2}{b(\r)\,\left(1+b(\r_I)\Omega^2\int_{\r_I}^{\r}\di y\,\frac{1}{b(y)}\right)}\,.
\eeqs

\subsection{Asymptotic analysis}

An immediate question we need to ask is  what are the (UV) divergences of the resulting four-dimensional theory.
We analyze the $\r_U$-dependence of the $\pi(q^2)$ functions, and impose conditions such that the resulting $\pi(q^2)$
behave in the same way as expected from sensible four-dimensional field theories.
As a first requirement, we must find that $\pi(0)=-b(\r_U)\gamma_0(\r_U)$ is finite when $\r_U\rightarrow+\infty$,
hence ensuring the EWSB is triggered by the strong-coupling dynamics in the dual field theory.
The second requirement is that $\pi^{\prime}(0)=-b(\r_U)\gamma_1(\r_U)$ must diverge, which is related to the fact  that
the strength of the (weak) gauging of the global symmetry is a free parameter. Alternatively, one can think of this in terms of the fact that the
zero modes of the bulk gauge bosons should be non-normalizable, in the absence of UV-localized boundary counter-terms. 
The third requirement is that  all the higher-order $\pi^{(n)}(0)$ must be finite, in order  to 
retain  predictive power and renormalizability of the dual weakly-coupled $U(1)^2$ gauge theory.
Finally, it must be the case that the counter-term that is needed to renormalize the vectorial propagator also renormalizes the axial-vectorial one, so that quantities such as the $S$ parameter are calculable.

We start with a model-independent simplified analysis. Let us assume that for asymptotically large $\r$, we have
\beqs
a(\r)&\propto&e^{\alpha\,\r}\,,\\
b(\r)&\propto&e^{\beta\,\r}\,.
\eeqs
The vectorial components behave as
\beqs
b(\r)\gamma_0(\r)&=&0\,,\\
b(\r)\gamma_1(\r)&\propto&\left\{\begin{array}{cc}
e^{\alpha\,\rho}\,+\,{\rm finite}\,,&\alpha\neq 0\cr
\r\,+\,{\rm finite}\,,\,&\alpha=0\end{array}\right.\,.
\eeqs
If $\alpha>0$, then for any $\beta$ there will be a $n$ such that
$b \gamma_n \propto \int e^{(n\alpha-\beta)\r}$ diverges, which implies that we cannot allow for this possibility.
If $\alpha<0$, then we would find that $b\gamma_1$ is UV-finite, which we cannot allow for either.
We are hence forced to set $\alpha=0$. In this case, we find in general that 
$b\gamma_n\propto \int\di\rho\,e^{-\beta\r}\r^n$, and this means that we must require that
$\beta>0$.
In summary, we require that $\alpha=0$ and $\beta>0$.

As an example,  in the case in which the original gravity background is AdS$_5$,
then $\delta=0=\theta$, and $\chi=0$, so that $A=\r$, and one finds that
\beqs
a(\r)&\propto&1\,,\\
b(\r)&\propto&e^{2\r}\,,
\eeqs
or equivalently $\alpha=0$, $\beta=2$, satisfying  all of the above requirements.

The axial-vector case is more subtle and model-dependent.
The zero-momentum $b \gamma_0$ depends on $\int_{\r_I}^{\r}\di y/b(y)$,
which converges in the UV provided $b\sim e^{\beta\r}$ with $\beta>0$.
Yet, a possible difficulty arises in this case in the IR, if $b(0)=0$.

Assume that near the IR end-of-space one has $b(\r)\simeq \r^B$.
One then finds that
\beqs
b(\r)\gamma_0(\r)&=&\frac{\r_I^{B}\Omega^2}{1+\r_I^B\Omega^2\int_{\r_I}^{\r}\di y/b(y)}
\,=\,\frac{\r_I^{B}\Omega^2}{1+\r_I^B\Omega^2\left({\cal O}(\r_I^{1-B})+{\rm finite}\right)}\,.\nonumber
\eeqs
We have two possibilities.
If $B<1$, the integral is IR-convergent, in which case we can ensure that the final result 
be finite and non-zero by redefining $\Omega^2\equiv \kappa\tilde{\Omega}^2/b(\r_I)$
and  by taking the $\r_I\rightarrow 0$ limit for fixed $\tilde{\Omega}$.
If $B > 1$, the integral is IR-divergent and we have
\beqs
\nonumber
b(\r)\gamma_0(\r)&\sim&\frac{\r_I^B\Omega^2}{1+{\cal O}(\r_I \Omega^2)}\,,
\eeqs
while for $B=1$ there is a corresponding logarithmic divergence. This means that there is no simple way to make this finite and non-vanishing, unless one accepts the idea of introducing
an IR-subtraction process similar to what done for the kinetic terms in relation to the UV divergences.
This is not compatible with an interpretation in terms of a dual field theory within the context of gauge-gravity duality.
In conclusion, we find that the absence of IR divergences, combined with a non-trivial result for EWSB,
 amounts to requiring $B<1$.

Finally, we fix the normalization of the gauge bosons by choosing 
\beqs
\frac{1}{e^2}&\equiv&-b(\r_U)\left(\frac{}{}\gamma_{V\,,\,1}(\r_U)+D\right)\,,
\eeqs
and hence we have
\beqs
\hat{S}&\simeq&\cos^2\theta_W\left(1-e^2\,\pi^{\prime}_A(0)\frac{}{}\right)\,=\,
e^2\cos^2{\theta_W}\,b(\r_U)\left(\frac{}{}\gamma_{A\,,\,1}(\r_U)-\gamma_{V\,,\,1}(\r_U)\right)\,.
\eeqs

In the specific case of our model, by comparing the two actions in Eq.~(\ref{Eq:S5}) and Eq.~(\ref{Eq:Seff4}),
we find that
\beqs
a(\r)&=&\kappa\,e^{(\delta+\omega)\chi}\,,\\
b(\r)&=&\kappa\,e^{2A+(\omega-\delta)\chi}\,.
\eeqs
The arguments discussed earlier imply that we must impose $\omega=-\delta$.
We use this choice in general, for all our background solutions, in the rest of this section.

\subsection{HSV solutions}

In the case of the HSV solutions, we have
\beqs
a(\r)&=&\kappa\,,\\
b(\r)&=&\kappa \,e^{2A-2\delta\,\chi}\,=\,\kappa \,e^{\frac{2}{4-\theta}\r}\,,
\eeqs
where we conveniently wrote:
\beqs
A&=&\frac{(\theta-3)}{3(\theta-4)}\,\r\,,\\
\chi&=&\frac{\theta}{\theta-4}\sqrt{\frac{\theta-3}{6\theta}}\,\r\,.
\eeqs
Notice that one might interpret this in terms of the fact that the background ultimately is always the same,
by reabsorbing $\r/(4-\theta)\rightarrow \r$. However, this rescaling would have a non-trivial effect on the boundary terms
$\Omega$ and $\kappa$, and hence this is not going to be apparently visible in the results of the analysis. Alternatively, we rescale also the boundary terms $\Omega$ and $\kappa$. We do not do
this because it would make it more difficult to compare to the confining solutions, which are 
physically more interesting.

In the present case, there is no end-of-space to the geometry,
and we can set $\r_I=0$, which also implies that $b(\r_I)=\kappa$ and hence $\Omega^2=\tilde{\Omega}^2$,
so that we can rewrite
\beqs
b(\r) \gamma_{V\,,\,0}(\r)&=&0\,,\\
b(\r) \gamma_{V\,,\,1}(\r)&=&-\kappa \r\,,\\
b(\r) \gamma_{A\,,\,0}(\r)&=&\frac{\kappa \Omega^2}{1-\frac{4-\theta}{2}\Omega^2\left(e^{-\frac{2}{4-\theta}\r}-1\right)}\,,\\
b(\r) \gamma_{A\,,\,1}(\r)&=&
-\frac{\kappa  \left((\theta -4)^2 \Omega ^2 \left(e^{\frac{2 \r}{\theta -4}}-1\right)
   \left((\theta -4) \Omega ^2 \left(e^{\frac{2 \r}{\theta -4}}-3\right)+8\right)+4 \r
   \left((\theta -4) \Omega ^2-2\right)^2\right)}{4 \left((\theta -4) \Omega ^2
   \left(e^{\frac{2 \r}{\theta -4}}-1\right)+2\right)^2}\,.
\eeqs
The last expression needs to be commented upon.
First of all, notice how for $\Omega=0$, one recovers $b(\r) \gamma_{A\,,\,1}(\r)=b(\r) \gamma_{V\,,\,1}(\r)=-\kappa \r$,
as expected.
As long as $\theta<4$, taking $\r\rightarrow+\infty$ and expanding for small $\Omega$
yields
\beqs
b(\r) \gamma_{A\,,\,0}(\r)&\rightarrow&\kappa\Omega^2\,,\\
b(\r) \gamma_{A\,,\,1}(\r)&\rightarrow&\kappa\left(-\r+\frac{1}{2}(\theta-4)^2\Omega^2\right)\,,
\eeqs
where we assumed $\r\gg0$. 
As a result, the physical parameters are
\beqs
M_Z^2&\simeq&-\frac{\pi_A(0)}{\pi^{\prime}_A(0)}\,=\,\frac{e^2\kappa \Omega^2}{1-\frac{e^2}{2}(\theta-4)^2 \kappa \Omega^2}\,
\simeq\,e^2 \kappa\Omega^2\,,\\
\hat{S}&\simeq&\frac{e^2}{2}\cos^2\theta_W\,(\theta-4)^2\kappa\Omega^2\,\simeq\,\frac{(\theta-4)^2}{2}\cos^2\theta_W\,M_Z^2\,.
\eeqs
Conversely, for $\theta>4$ all the precision parameters, including $\hat{S}$, diverge in the $\r\rightarrow +\infty$ limit,
and hence the precision parameters are not calculable in this case.
This is a general result, valid also for all the other classes of models we consider: for $\theta>4$ 
we cannot recover results that can be 
interpreted in terms of a renormalizable, sensible four-dimensional field theory.

In order to make phenomenological sense of the $\theta$ dependence, in a context in which we have several other parameters
entering non-trivially our final results, we make use of the observation that (provided $\Omega^2$ is small),
we expect $\hat{S}\propto M_Z^2/M_{\r}^2$, where $M_{\rho}$ is the first massive excitation of the vector 
modes in the four-dimensional theory (the techni-rho meson).

The solution to the bulk equations and boundary conditions for 
the vectorial modes  read:
\beqs
\gamma_V(q,\r)&=&
\frac{q \,e^{\frac{\r}{\theta -4}} \left(J_0[q (\theta -4)] Y_0\left[e^{\frac{\r}{\theta
   -4}} q (4-\theta )\right]-Y_0[q (4-\theta)] J_0\left[e^{\frac{\r}{\theta -4}} q
   (\theta -4)\right]\right)}{Y_0[q (4-\theta )] \left(-J_1\left[e^{\frac{\r}{\theta
   -4}} q (\theta -4)\right]\right)-J_0[q (\theta -4)] Y_1\left[e^{\frac{\r}{\theta
   -4}} q (4-\theta)\right]}\,,
\eeqs
where it is explicitly visible that the IR boundary condition is satisfied. Notice a technical fact:
 numerically one has to solve the original linear, second-order equation for $v(q,r)$, because of the poles in $\gamma$.

Adding the counter-term and taking $\r\rightarrow +\infty$, a finite answer can be recovered (provided $\theta<4$): 
\beqs
\pi_V(q^2)&=&\frac{1}{e^2} q^2+q^2 (\theta-4)\kappa\left[\gamma_E-\frac{\pi Y_0[(4-\theta)q]}{2\,J_0[(\theta-4)q]}
+\log\left(\frac{(4-\theta)q}{2}\right)\right]\,.
\eeqs

The spectrum can be read off the zeros of $\pi_V(q^2)$.
Notice that the poles of this expression are fixed by the zeros of the Bessel $J_0[(\theta-4)q]$, which are
for $q(4-\theta)=2.4, 5.5, 8.6,\cdots$.
Because $e=\sqrt{4\pi\alpha}$ is fixed experimentally, the precise position of the zeros, and in particular of the first 
resonance, depends on $\kappa$.
For small $\kappa$, then $M_{\r}\simeq \frac{2.4}{4-\theta}$, just above the first pole,
while for large values of $\kappa$ the mass is higher, but cannot come close to the next pole: $M_{\r}\simeq \frac{4.6 }{4-\theta}$.
In conclusion then, depending of $\kappa$ and for $\theta<4$:
\beqs
2.4&<&(4-\theta)\,M_{\r}\,<\,4.6\,.
\eeqs

The exact results for the axial-vector case, which depend
on $\theta$ as well as $\Omega$, is
\beqs
&&\gamma_A(q,\r)\,=\,q e^{\frac{\r}{\theta -4}} \,\times\,\\
&&\frac{\left(q J_0[q (\theta -4)]-\Omega ^2 J_1[q
   (\theta -4)]\right) Y_0\left[e^{\frac{\r}{\theta -4}} q (4-\theta)\right]-\left(q
   Y_0[q (4-\theta)]+ \Omega ^2 Y_1[q (4-\theta)]\right)
   J_0\left[e^{\frac{\r}{\theta -4}} q (\theta -4)\right]}{ \left(\Omega
   ^2 Y_1[q (4-\theta )]+q Y_0[q (4-\theta )]\right)\left(\frac{}{}- J_1\left[e^{\frac{\r}{\theta
   -4}} q (\theta -4)\right]\right)-\left(-\Omega ^2 J_1[q (\theta -4)]+q J_0[q (\theta
   -4)]\right) Y_1\left[e^{\frac{\r}{\theta -4}} q (4-\theta )\right]}\,.\nonumber
\eeqs
This can be obtained from the vector case, provided we apply the replacements
\beqs
J_0[q (\theta -4)]&\rightarrow&J_0[q (\theta -4)]-\frac{\Omega^2}{q} J_1[q
   (\theta -4)]
\,,\\
Y_0[q (4-\theta )]&\rightarrow&Y_0[q (4-\theta)]+\frac{\Omega^2}{q} Y_1[q
   (4-\theta )]\,.
\eeqs
Armed of this additional observation, we conclude that
\beqs
\pi_A(q^2)&=&\frac{1}{e^2} q^2+q^2 (\theta-4)\kappa\left[\gamma_E-\frac{\pi \left(Y_0[q (4-\theta)]+\frac{\Omega^2}{q} Y_1[q
   (4-\theta )]\right)}
{2\,\left(J_0[q (\theta -4)]-\frac{\Omega^2}{q} J_1[q
   (\theta -4)]\right)}
+\log\left(\frac{(4-\theta)q}{2}\right)\right]\,.
\eeqs

\begin{figure}[h,t]
\begin{center}
\begin{picture}(220,150)
\put(20,5){\includegraphics[height=5cm]{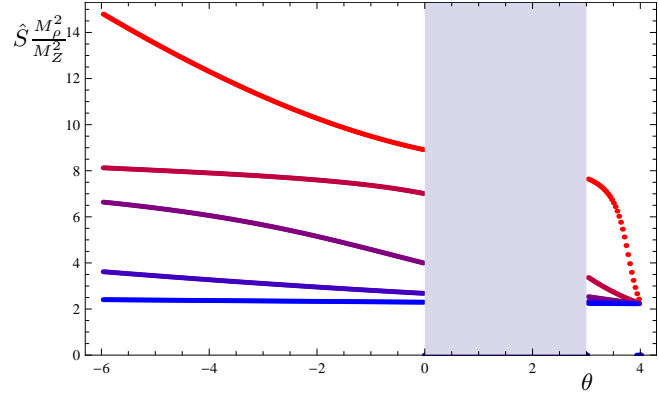}}
\put(0,130){$\hat{S}\frac{M_{\r}^2}{M_{Z}^2}$}
\put(215,0){$\theta$}
\end{picture}
\caption{The combination $\hat{S}M_\r^2/M_Z^2$ as a function of $\theta$ for the HSV backgrounds. The other parameters are
$e=\sqrt{4\pi\alpha}$,  $\Omega=1/25$, and $\kappa=(20,3,1,1/3,1/20)$ from top (red) to bottom (blue).
For $\hat{S}$ we used the approximation valid for small $\Omega$, while the masses of $Z$ and $\rho$ are
obtained numerically from the exact form of $\pi_V$ and $\pi_A$ in the main text. The shaded region does not correspond to physical backgrounds.}
\label{Fig:Shyperscaling}
\end{center}
\end{figure}

To first approximation the
product $\hat{S} M_{\r}^2/M_Z^2$ is independent of $\theta$, at least for small $\kappa$ and $\Omega^2$,
and one finds that $\hat{S}M_{\r}^2/M_Z^2\simeq 2.2$.
This is illustrated with the numerical results in Fig.~\ref{Fig:Shyperscaling}.
Irrespectively of $\kappa$ and $\Omega$,
we find the bound
\beqs
M_{\rho}&\gsim&M_Z\sqrt{\frac{2.2}{\hat{S}}}\simeq\,2.4\,\,{\rm TeV}\,,
\eeqs
where we imposed the bound $\hat{S}_{max} \sim 0.003$.
By making use of the $1\sigma$ bound for light Higgs $\hat{S}_{max} \sim 0.0013$, this bound raises to $M_{\rho}>3.7$ TeV.
This is consistent with the typical bounds obtained from models based on AdS five-dimensional backgrounds~\cite{TCholobottom}. The parameter $\kappa$ is expected to scale as $N$ in the large-$N$ limit of the dual field theory.

\subsection{Confining solutions}

In the case of the confining solutions, all of which are defined for $\r>0$, we have
\beqs
a(\r)&=&\kappa\,,\\
b(\r)&=&2^{\frac{2 (\theta -3)}{3 (\theta -4)}} \kappa  
\sinh ^{\frac{\sqrt{(\theta -3) \theta   }-2}{-4+\theta }}\left(\frac{\r}{2}\right) 
\cosh ^{\frac{\sqrt{(\theta -3) \theta   }+2}{-\theta +4}}\left(\frac{\r}{2}\right)\,,
\eeqs
By construction, the first two contributions to $b(\r)\gamma_V(q,\r)$ are the same as for the hyperscaling case:
\beqs
b(\r) \gamma_{V\,,\,0}(\r)&=&0\,,\\
b(\r) \gamma_{V\,,\,1}(\r)&=&-\kappa \r\,,
\eeqs
and this also means that the counter-term is the same
\beqs
D\,b(\r_U)&=&-b(\r_U)\gamma_{V\,,\,1}(\r_U)\,-\,\frac{1}{e^2}\,, \nonumber
\eeqs
rather unsurprisingly, given that the analysis of the counter-terms depends only on the large-$\r$ asymptotics.

By expanding for small $\r$, we find that
\beqs
b(\r)&\propto& \r^{\frac{-2+\sqrt{\theta(\theta-3)}}{\theta-4}}\,.
\eeqs
As long as $\theta<0$, the exponent fulfills the requirement $B<1$. However, when $\theta>3$
the exponent yields $B>1$. 
Hence, we are forced to abandon the idea of modeling EWSB in terms of just the IR boundary conditions
for the bulk gauge bosons in the cases where $\theta>3$. More complicated possibilities exist, such as 
allowing for a new bulk scalar to introduce EWSB, but this goes beyond the purposes of this study.

In the alternative background, the functions $a$ and $b$ are
\beqs
a(\r)&=&\kappa\,,\\
b(\r)&=&2^{\frac{2 (\theta -3)}{3 (\theta -4)}} \kappa  
\sinh ^{\frac{\sqrt{(\theta -3) \theta   }+2}{4-\theta }}\left(\frac{\r}{2}\right)
\cosh ^{\frac{\sqrt{(\theta -3) \theta  }-2}{\theta -4}}\left(\frac{\r}{2}\right)\,,
\eeqs
  which when expanded for small-$\r$ yields
  \beqs
  b(\r)&\propto&\r^{\frac{\sqrt{(\theta -3) \theta   }+2}{4-\theta }}\,.
  \eeqs
  Again, for $\theta<0$ this yields an IR-convergent result for the integral entering $b \gamma_{0\,A}(\r)$.

\begin{figure}[h,t]
\begin{center}
\begin{picture}(220,150)
\put(20,5){\includegraphics[height=5cm]{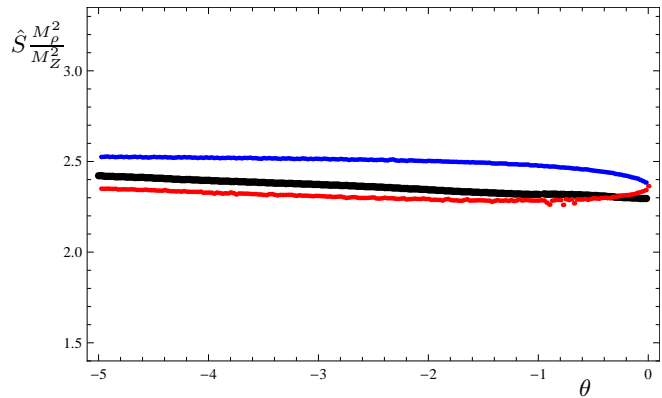}}
\put(0,130){$\hat{S}\frac{M_{\r}^2}{M_{Z}^2}$}
\put(215,0){$\theta$}
\end{picture}
\caption{The combination $\hat{S}M_\r^2/M_Z^2$ as a function of $\theta$. The other parameters are
$e=\sqrt{4\pi\alpha}$,  $\sqrt{\kappa}\tilde{\Omega}=1/25$ and $\kappa=1/20$, and we used
$\r_I=10^{-6}$, $\r_U=22/|4-\theta|/9$.
All the results are obtained from the numerical solutions. Shown are the hyperscaling case (black), 
 the confining case (red) and the alternative confining case (blue).}
\label{Fig:numerics}
\end{center}
\end{figure}

   We performed a numerical study of the precision parameter $\hat{S}$, in relation to $M_{\r}^2$ and $M_Z^2$,
   for the confining case, comparing the results to the HSV case.
   In Fig.~\ref{Fig:numerics}  we show the comparison between the 
   results of the numerical study.
   At least for $\theta<0$, the confining solutions yield almost the same result for the combination $\hat{S}M_{\r}^2/M_Z^2$ 
   as the HSV ones, up to small effects.
  The conclusion of this exercise being that the HSV and confining backgrounds give similar results for $\theta<0$, at least
  in the regime in which $\kappa$ is not large. For completeness we performed this calculation also for the alternative solution, in spite of its pathological character, and we find that the result for the $S$ parameter is close to the HSV and confining ones.

\section{Discussion}

We start this discussion section by summarizing the main results of our study.
There are three main classes of general messages we can take away from the study of the toy models in this paper,
which we group by theme:
\begin{itemize}
\item results related to the breaking of scale invariance and the appearance of a light dilaton in the scalar spectrum of the (putative) dual field theory,
\item results related to modeling EWSB, precision physics and the properties of the excited spin-1 resonances of the dual field theory,
\item results related to the formal properties of backgrounds that are 
not approximately AdS, particularly in the HSV case.
\end{itemize}

Barring fine-tuning of the UV boundary conditions~\cite{EFHMP}, there exist two 
regions in the space of $\theta$ for which a parametrically light scalar emerges.
The first one is the case where $-1\ll\theta < 0$.
In this region, the metric is everywhere close to AdS$_5$, and hence the dual field theory is always close to being scale invariant.
The fact that a light scalar state emerges from the calculation of the mass spectrum is true both in the case of HSV backgrounds 
as well as the semi-realistic backgrounds with an end-of-space, that we called confining.

In order to gauge how realistic it is to obtain a light scalar by building a model with these  values of $\theta$,
we recall the fact  that the case $\theta=-1$ corresponds to a physical case
that can be uplifted to a  full string-theory background. This model results from the compactification on a circle
of one of the external dimensions of the AdS$_6\times S^4$ solution of massive type-IIA~\cite{ADS6S4,ADS6,F4QCD4,EFHMP}.
The spectrum of the dual theory does not contain a parametrically light state.
Compactifying on larger manifolds in general yields values for $\theta$ further from zero (examples  in~\cite{Perlmutter} 
yield $\theta=-2,-1,7,8,9$).

In the HSV case  we found a parametrically light scalar also for $3\leq \theta \ll 4$.
$\theta=3$ is the hyperscaling coefficient of flat space in five dimensions.
While the interpretation of the boundary theory in terms of a 
lower dimensional field theory is still a matter of ongoing studies, in view of this result it is tempting to interpret it in
terms of  spontaneously broken scale invariance.
But the parametrically light state we found in the HSV case disappears for the more general and semi-realistic confining solutions.
We do not know whether this is true under more general conditions, yet it suggests that the simple
modeling of confinement  we adopted here cannot yield a parametrically light state for $3<\theta<4$.

The general message is that in order for a parametrically light scalar state to be present in the spectrum of the dual 
field theory, one needs to find a  non-trivial model, in which either the gravity background is always very close to AdS$_5$, or to 
flat space, and in the latter case the mechanism for confinement is 
 going to involve a combination of dynamical scalars in the five-dimensional $\sigma$-model,
 not captured by the toy models of this paper.

Regarding EWSB,  we made a drastic (though quite common in the literature) simplifying assumption in this study: we assumed 
that the bulk (five-dimensional) equations for the vector and axial-vector  bosons be identical, and only IR boundary 
conditions are responsible for symmetry breaking, with no further degrees of freedom involved in the dynamics.
Within this framework, we found a  bound on the mass of the lightest
spin-1 excitation (the techni-$\r$ in the dual theory)
compatible with precision physics. Adopting  the $1\sigma$ bounds on $\hat{S}$ for a light Higgs implies that for all the models discussed in this paper $M_{\r}>3.7$ TeV, 
(down to $2.4$ TeV if one uses $3\sigma$ bounds on $\hat{S}$)~\cite{Barbieri}.
This bound does not depend on the value of $\theta$, the hyperscaling coefficient. The bound becomes more stringent 
for large values of the parameter $\kappa$,
which is related to the effective strength of the self-couplings of the heavy resonances.
And the bound  depends only marginally  on whether one models confinement with
a hard-wall in the IR or with a dynamical end-of-space.

At a more formal level, there is a substantial difference between what happens in the four intervals $\theta<0$, $0<\theta<3$,
$3<\theta<4$ and $\theta>4$.
In the case $\theta<0$ we have been able to perform all the calculations one can perform in AdS, without major difficulty.
Aside from the fact that because the space is not asymptotically AdS
the calculation of observables related to EWSB enforced a very special choice
for the conformal factor appearing in the five-dimensional action in front of the kinetic term for the gauge bosons.
We do not have a reason of principle (nor a dynamical explanation) for why such a factor should be there, except from the pragmatic one that 
this is the only way to ensure that the UV-asymptotic behavior of the two-point functions reproduces the main properties 
expected from a four-dimensional gauge theory.

The case $0<\theta<3$ is known to yield  violation of the null  energy condition~\cite{Dong:2012se}.
Interestingly, in the HSV case we could extend explicitly the equations for the scalar fluctuations
in this region, and we found a tachyonic state in the spectrum, confirming the pathological character of this regime.

The region $3<\theta<4$ yields a surprise: the appearance of IR-divergences prevents the implementation of 
EWSB (or chiral symmetry breaking) purely by IR boundary conditions. 
This implies that if any model is found in the top-down approach to holography 
in which the background exhibits hyperscaling violation with coefficient in this range, then not only confinement (as we said speaking about the spectrum of scalar states), but also chiral symmetry breaking will require some non-trivial amount of work in order to be implemented successfully.
It is clear that the deep-IR dynamics of a realistic model will have to deviate quite substantially from the simplified treatment we adopted 
in the present work, at least in this interval of possible values for $\theta$.

Finally, the region $\theta>4$ looks unlikely to be of much relevance for phenomenological purposes.
Besides the fact that it is not going to yield a parametrically light scalar state,
it appears impossible to implement a calculable model of EWSB in this case, because of the bad UV-divergent behavior of the non-polynomial $q^2$ 
dependence of the 2-point correlation functions, which ultimately requires the introduction of an infinite number of independent local counter-terms 
(UV-boundary localized terms)
in order to cancel all the independent divergences.

We now turn to the questions we highlighted in the introduction, and comment on how much what we did helps
the general goals of the physics program of exploration of realistic models of DEWSB via holography.
The first question regards the systematics underlying the modeling of holographic systems that yield 
a parametrically light scalar state in the spectrum.
The appearance of hyperscaling violation per se does not yield this dynamical feature.
Only when the hyperscaling coefficient is close to 0 or 3 does a light scalar state appear in the spectrum.
It is not surprising that for $\theta\simeq 0$ we should find a light dilaton, and yet it is in general quite difficult to build models
from the top-down with this feature. 

Most interesting from the model-building perspective is that for $\theta\simeq 3$, when the metric is close to flat space,
the possibility of a light scalar emerges.
However, if this were to emerge dynamically, the deep-IR of the system cannot be simple: we saw that in the 
one-scalar sigma-model, with exponential potential, considering backgrounds that have an end-of-space
effectively removes the parametrically light scalar and gives it a non-trivial mass.

The second question we highlighted in the introduction had to do with the fact that while there are 
a small number of top-down models in which a parametrically light scalar state appears in the spectrum, their geometries are not close to AdS, and little is known about such case in the more general bottom-up approach
(with the possible exception of~\cite{Barcelona}). In this paper we explored a broad class of bottom-up models, showing that there is at play a combination of general as well as model-dependent factors. Top-down models allow for a richer dynamics, the many subtleties of which are not fully captured by these bottom-up models.

Finally, the last questions we posed reflected the fact that in top-down models the complication of the background means that it is 
hard to use them to build realistic EWSB models.
We limited our attention to the very simplest possible way to implement EWSB, namely via IR boundary conditions.
We found a general result: suppressing the $S$-parameter within the experimental bounds requires that the 
mass of the lightest spin-1 excitation of the model must be in the multi-TeV range.
This is in good agreement with expectations from generic EFT arguments.

In conclusion, we constructed and discussed large classes of five-dimensional backgrounds 
exhibiting hyperscaling violation, in the context of the bottom-up approach to holography.
We systematically studied the scalar spectrum, and found regions of parameter space in which one
state is parametrically light.
We used the resulting toy models to describe the dynamics leading to EWSB,
which we modeled by the choice of IR boundary conditions.
We found that  precision electroweak constraints imply that the mass of the first spin-1 excited vector state 
is in the range of few TeV, beyond accessibility for the first run of LHC, but testable in the future LHC program.

\begin{acknowledgments}
\end{acknowledgments}

The work of MP is supported in part by the STFC Consolidated Grant ST/L000369/1.
RL is supported by the STFC Doctoral Training Grant ST/I506037/1. DE is supported by the NSF CAREER Award PHY-0952630 and by the DOE through grant DE-SC0007884.


\vspace{1cm}

\appendix

\section{Equations of motion and boundary conditions for the scalar fluctuations}

We report here explicitly the bulk equations of motion as well as boundary conditions for the scalar fluctuations in the confining backgrounds given by Eq.~\eqref{eq:confiningbackgroundA} and Eq.~\eqref{eq:confiningbackgroundchi}:
\beqs
	\mathfrak{a}'' + \coth(\rho) \mathfrak{a}' + \frac{6(\theta - 4)}{\theta + 4\cosh(\rho) \left( (\theta-3) \cosh(\rho) + \theta \sqrt{\frac{\theta-3}{\theta}} \right)} \mathfrak{a} + e^{2\delta \chi -2A} q^2 \mathfrak{a} &=&0, \\
	\frac{2(\theta -3)^2 \left( 2 + \sqrt{\frac{\theta}{\theta-3}} \cosh(\rho) \right)^2}{3(\theta-4) \sinh(\rho) \left( 2(\theta-3) \cosh(\rho) + \theta \sqrt{\frac{\theta-3}{\theta}} \right)} \mathfrak{a}' + \frac{2(\theta-3) \left( 2 +  \sqrt{\frac{\theta}{\theta-3}} \cosh(\rho) \right)}{\theta + 4\cosh(\rho) \left( (\theta-3) \cosh(\rho) + \theta \sqrt{\frac{\theta-3}{\theta}} \right)} \mathfrak{a} + \\ e^{2\delta \chi -2A} q^2 \mathfrak{a} \Bigg|_{\rho_i} &=& 0, \nonumber
\eeqs
where
\SP{
	e^{2\delta \chi -2A} = 
  2^{-\frac{2 (\theta -3)}{3 (\theta -4)}} \sinh
   \left(\frac{\r}{2}\right)^{\frac{\theta +1}{\sqrt{\frac{\theta -3}{\theta }}
   \theta -2}} \cosh \left(\frac{\r}{2}\right)^{-\frac{\theta
   +1}{\sqrt{\frac{\theta -3}{\theta }} \theta +2}}.
}

\section{IR and UV regulators}

The numerical calculation of the mass spectrum of scalar fluctuations has been performed by introducing two
regulators, $\r_I$ and $\r_U$. This amounts to the introduction into the problem of two new, spurious, unphysical scales.
We performed an extensive study of the effect of such regulators, to ensure that 
the results we quote reproduce the physical ones (in which the regulators have been removed).
We report here only the results of this analysis for $3<\theta<4$, which proved to be the hardest numerically.

\begin{figure}[h,t]
\begin{center}
\begin{picture}(400,401)
\put(120,270){\includegraphics[height=4cm]{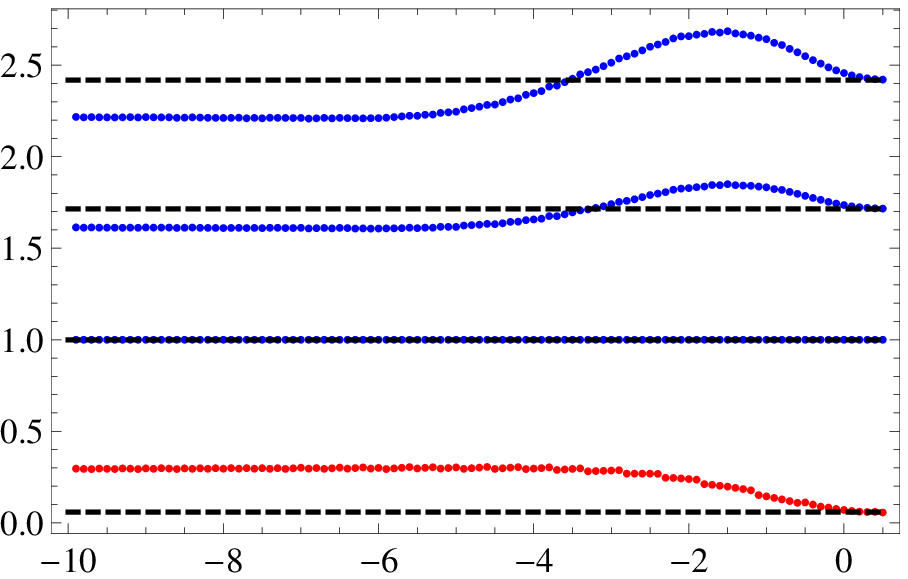}}
\put(20,135){\includegraphics[height=4cm]{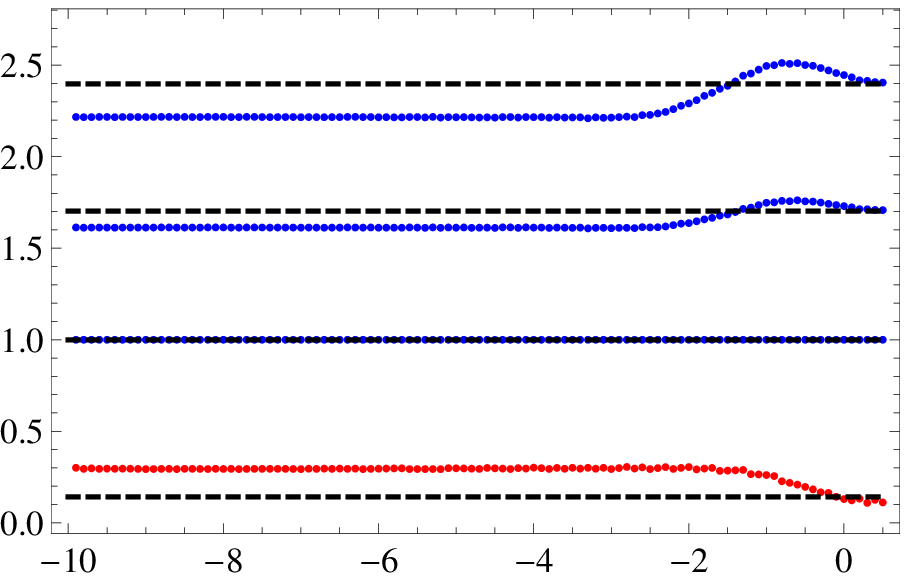}}
\put(220,135){\includegraphics[height=4cm]{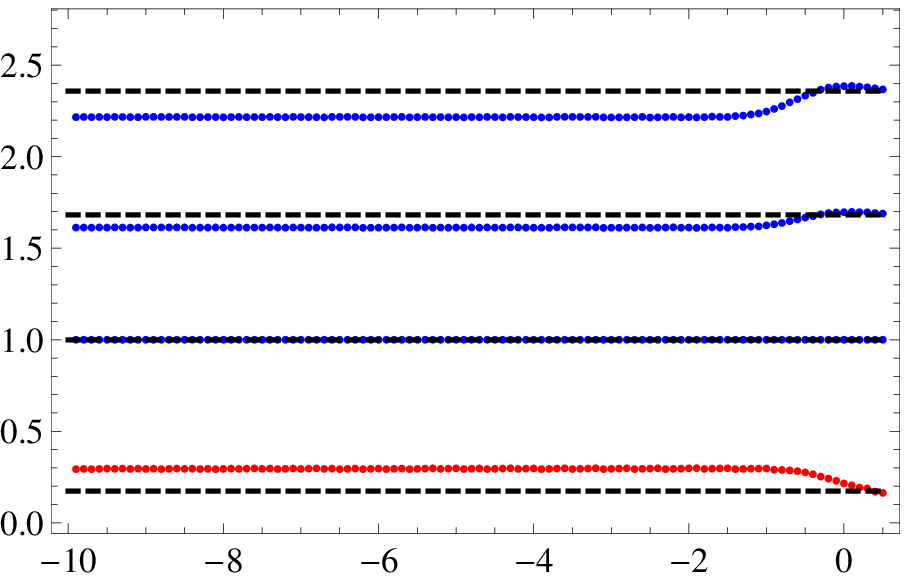}}
\put(20,5){\includegraphics[height=4cm]{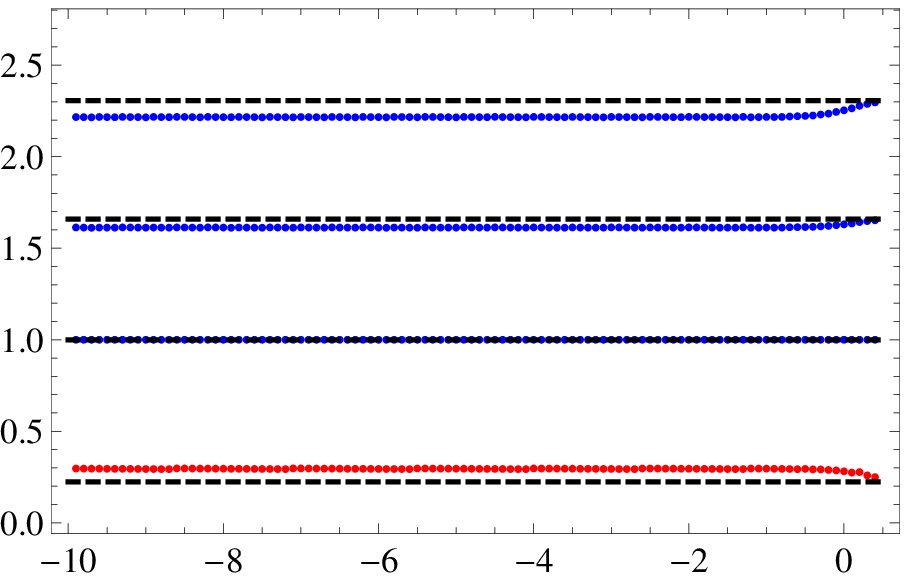}}
\put(220,5){\includegraphics[height=4cm]{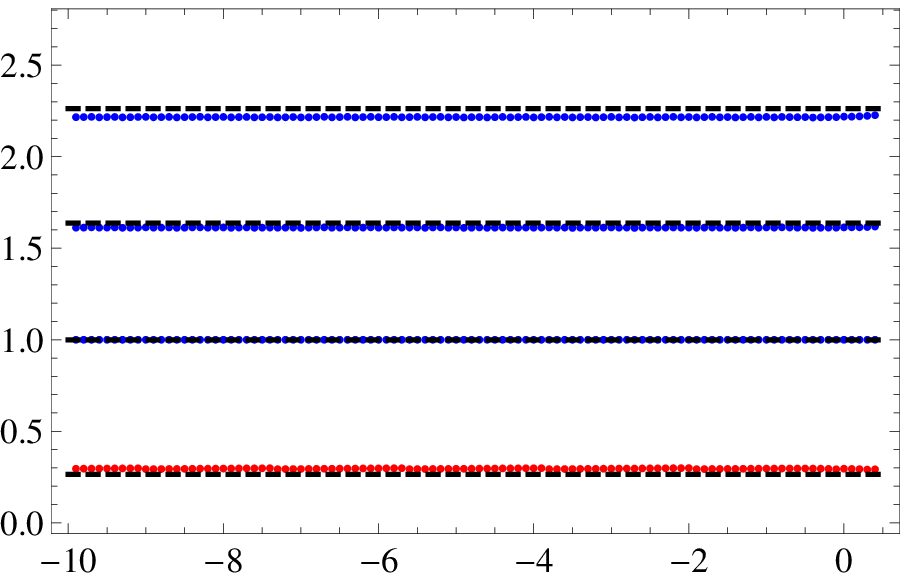}}
\put(110,375){$\frac{q}{q_1}$}
\put(210,240){$\frac{q}{q_1}$}
\put(210,105){$\frac{q}{q_1}$}
\put(10,240){$\frac{q}{q_1}$}
\put(10,105){$\frac{q}{q_1}$}
\put(330,0){$\log_{10}\rho_I-\log_{10}(4-\theta)$}
\put(330,130){$\log_{10}\rho_I-\log_{10}(4-\theta)$}
\put(130,0){$\log_{10}\rho_I-\log_{10}(4-\theta)$}
\put(130,130){$\log_{10}\rho_I-\log_{10}(4-\theta)$}
\put(230,265){$\log_{10}\rho_I-\log_{10}(4-\theta)$}
\end{picture}
\caption{The spectrum $q=\sqrt{q^2}$ of the scalar bound states obtained fixing $\r_U=\frac{10}{4-\theta}$,
and varying $\r_I$, for solutions with end-of-space at $\r_0=0$.
 The blue and red dots represent the spectra computed for solutions
that yield an end-of-space, while the black dots are obtained for the hyperscaling solutions.
The plot refer to the solutions with $\theta=3.031,3.101,3.251,3.501,3.751$ (left to right, top to bottom).
The masses are normalized to units in which the second state has $q_1^2=1$.}
\label{Fig:IR}
\end{center}
\end{figure}

In Figure~\ref{Fig:IR}, we show the scalar mass spectrum for the confining solutions
(blue and red dots) as a function of  $\r_I$  
for various choices of $\theta$, at fixed value of $\r_U$,
and compare to the spectrum obtained for  the HSV solutions (black dots).

When $\r_I$ is large, the spectrum agrees with the HSV case, 
because in this case one is putting the regulator at a scale larger than that set by the end-of-space.
When using small values of $\r_I$, the results do not depend on $\r_I$ itself,
which allows us to extrapolate them to the case in which the regulator has been removed.
In all the plots, one sees that this is the case, provided $\r_I<\r_I^{\ast}$, where 
$\log_{10} (\r_I^{\ast})-\log_{10}(4-\theta)\simeq 1+2\log_{10} (\theta-3)$
is an estimate of a fiducial value for the IR cutoff based on plots of this type.
In particular, this estimate highlights a technical difficulty: when $\theta$ is very close to the case $\theta=3$, if becomes
increasingly difficult to compute the spectrum, as this would require very high numerical precision and very small values of $\r_I$.

\begin{figure}[h,t]
\begin{center}
\begin{picture}(400,401)
\put(120,270){\includegraphics[height=4cm]{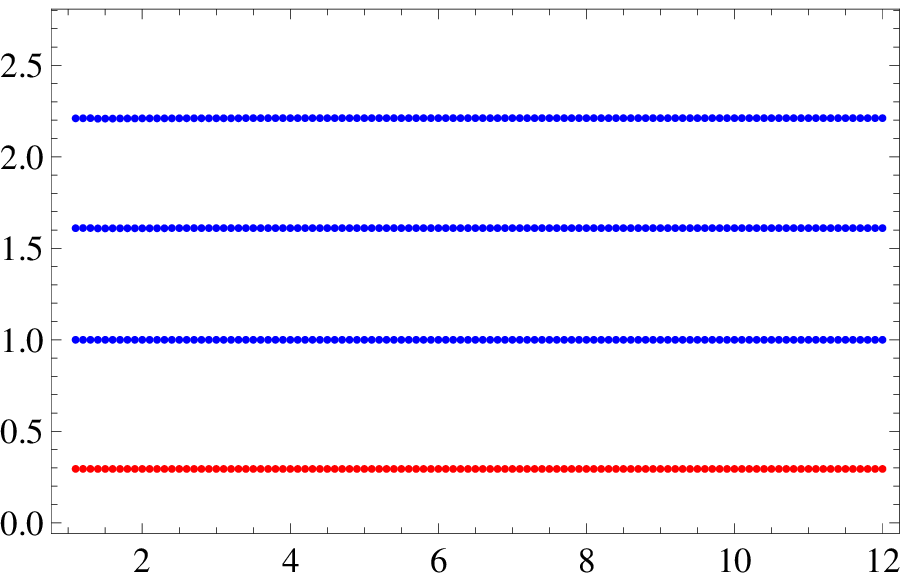}}
\put(20,135){\includegraphics[height=4cm]{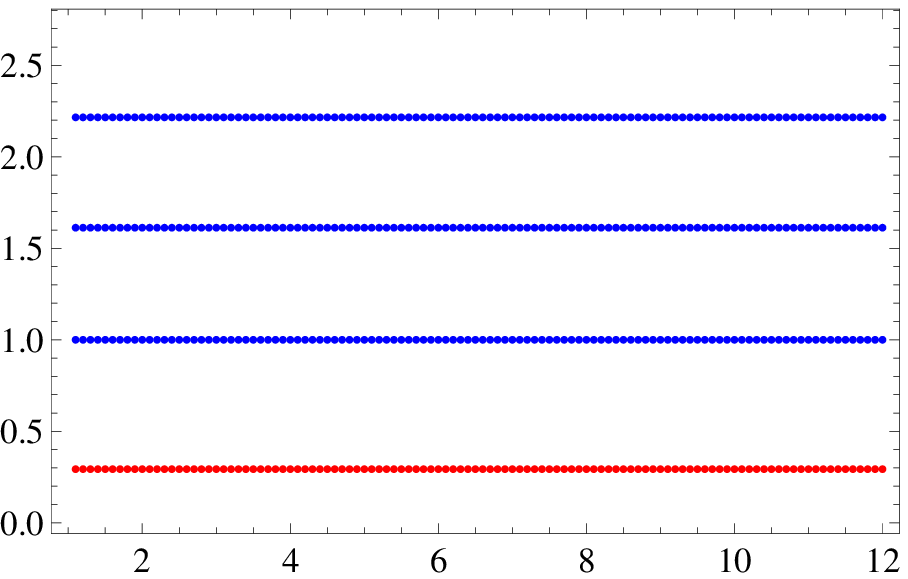}}
\put(220,135){\includegraphics[height=4cm]{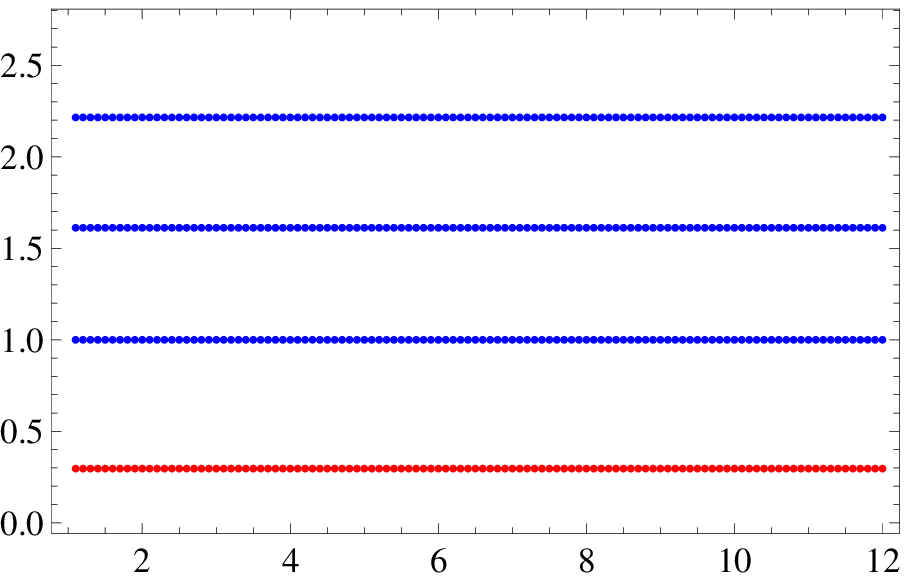}}
\put(20,5){\includegraphics[height=4cm]{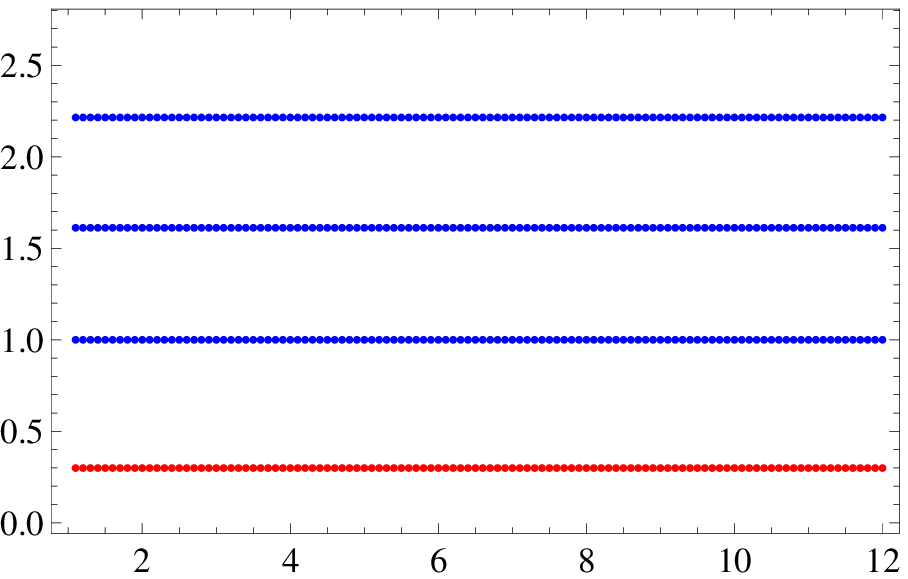}}
\put(220,5){\includegraphics[height=4cm]{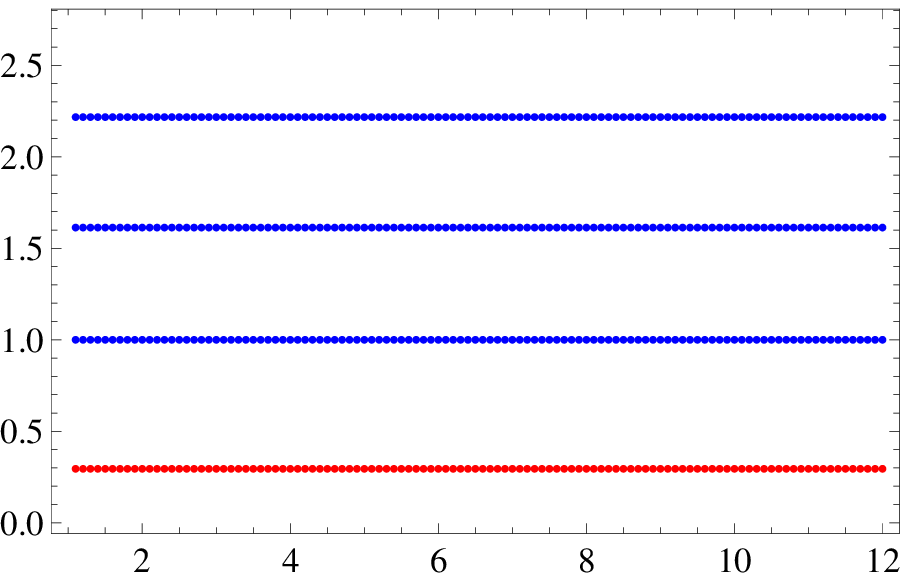}}
\put(110,375){$\frac{q}{q_1}$}
\put(210,240){$\frac{q}{q_1}$}
\put(210,105){$\frac{q}{q_1}$}
\put(10,240){$\frac{q}{q_1}$}
\put(10,105){$\frac{q}{q_1}$}
\put(330,0){$\rho_U/(4-\theta)$}
\put(330,130){$\rho_U/(4-\theta)$}
\put(130,0){$\rho_U/(4-\theta)$}
\put(130,130){$\rho_U/(4-\theta)$}
\put(230,265){$\rho_U/(4-\theta)$}
\end{picture}
\caption{The spectrum $q=\sqrt{q^2}$ of the scalar bound states obtained fixing $\log_{10}\r_I=\log_{10}(4-\theta)-7$,
and varying $\r_U$, for solutions with end-of-space at $\r_0=0$.
 The blue and red dots represent the spectra computed for solutions
that yield an end-of-space.
The plot refers to the solutions with $\theta=3.031,3.101,3.251,3.501,3.751$ (left to right, top to bottom).
The masses are normalized to units in which the second state has $q_1^2=1$.}
\label{Fig:UV}
\end{center}
\end{figure}

In Fig.~\ref{Fig:UV} we show the analogous study of the UV dependence.
As can be seen, the dependence on the UV cutoff is virtually invisible, within the choices of parameters we made.

\section{Conventions for the $S$ parameter}

We start with the standard definitions of  $\pi(q^2)$ in the natural basis of the $U(1)_L\times U(1)_R$ gauge bosons,
by setting the normalization conventions:
\beqs
\pi_{LL}(0)&=&-\frac{1}{4}g^2v_W^2\,,\\
{\pi}^{\prime}_{RR}(0)&=&
{\pi}^{\prime}_{LL}(0)\,=\,1\,,
\eeqs
and we also define $g^{\prime}\equiv g \tan\theta_W$. The precision parameter is defined as
\beqs
\hat{S}&\equiv&\frac{g}{g^{\prime}}\pi_{LR}^{\prime}(0)\,.
\label{Eq:Sdefinition}
\eeqs
We take as indicative experimental bound the $3\sigma$ limit $\hat{S}\lsim 0.003$~\cite{Barbieri}.

In order to go from the $(A_{\mu},V_{\mu})$ basis of Eq.~(\ref{Eq:Suseful}) to this  basis, 
we first redefine  both gauge bosons by the same 
rescaling, in order to make the vector canonically normalized $(A_{\mu}',V_{\mu}')=\frac{1}{g\sin\theta_W} (A_{\mu},V_{\mu})$.
Secondly we perform a rotation 
\beqs
L'_{\mu}&=&\cos\theta_W \, A_{\mu}'\,+\,\sin\theta_W\,V_{\mu}'\,,\\
R'_{\mu}&=&-\sin\theta_W \, A_{\mu}'\,+\,\cos\theta_W\,V_{\mu}'\,,
\eeqs
with $\tan\theta_W\equiv g^{\prime}/g$ the weak mixing angle relating the gauge couplings $g$ and $g^{\prime}$ of
the four-dimensional gauge bosons for $U(1)_L$ and $U(1)_R$, respectively.
Finally, we redefine the fields $(L_{\mu},R_{\mu})\equiv(\ell L'_{\mu}, {\rm r} R'_{\mu})$ so that the kinetic term becomes canonical.
The final result is the polarization in the $(L_{\mu},R_{\mu})$ basis:
\beqs
\pi&=&g^2\sin^2\theta_W\left(\begin{array}{cc}
\frac{1}{\ell} & 0\cr
0 & \frac{1}{\rm r}\end{array} \right)
\left(\begin{array}{cc}
\cos\theta_W & \sin\theta_W\cr
-\sin\theta_W & \cos\theta_W\end{array} \right)
\,\hat{\pi}\,
\left(\begin{array}{cc}
\cos\theta_W & -\sin\theta_W\cr
\sin\theta_W & \cos\theta_W\end{array} \right)
\left(\begin{array}{cc}
\frac{1}{\ell} & 0\cr
0 & \frac{1}{\rm r}\end{array} \right)\,.
\eeqs
By choosing 
\beqs
\ell&=&\sqrt{\sin^2\theta_W+g^2\sin^2\theta_W \pi_A^{\prime}(0)\cos^2\theta_W}\,,\\
{\rm r}&=&\sqrt{\cos^2\theta_W+g^2\sin^2\theta_W \pi_A^{\prime}(0)\sin^2\theta_W}\,,
\eeqs
and looking at the $q^2$-independent part of the $LL$ entry one finally learns that 
\beqs
M_Z^2&=&\frac{1}{4}v_W^2\,\frac{1+g^2\pi_A^{\prime}(0)\cos^2\theta_W}{\cos^2\theta_W \pi_A^{\prime}(0)}\,\
\simeq\,\frac{1}{4\cos^2\theta_W}g^2v_W^2\,,
\eeqs
where the approximation in exact for $\pi_A^{\prime}(0)=1/(g^2\sin^2\theta_W)$.

We finally have the expression for the $\hat{S}$ parameter we need:
\beqs
\hat{S}&=&
\frac{\cos (\theta_W) \left(\cot (\theta_W)-g^2 \pi_A^{\prime}(0) \sin \theta_W \cos\theta_W\right)}{\sqrt{\left(g^2 \pi_A^{\prime}(0)
   \cos ^2(\theta_W)+1\right) \left(g^2 \pi_A^{\prime}(0) \sin ^4(\theta_W)+\cos ^2(\theta_W)\right)}}\\
   &\simeq&\cos^2\theta_W\left(1-g^2\sin^2\theta_W \pi_A^{\prime}(0)\frac{}{}\right)\,,
\eeqs
where the last expression assumes that $\pi_A^{\prime}(0)\simeq1/(g^2\sin^2\theta_W)$, in agreement with the smallness of the 
$\hat{S}$ parameter required for phenomenological reasons. Notice that this agrees with the definition
$\hat{S}=\cos^2\theta_W(\hat{\pi}_V^{\prime}(0)-\hat{\pi}_A^{\prime}(0))$ used elsewhere in the literature, 
provided the normalizations
of the original fields are chosen so that $\hat{\pi}_V^{\prime}(0)=1$ instead of $\hat{\pi}_V^{\prime}(0)=1/(g^2\sin^2\theta_W)$.


\end{document}